\documentclass[journal, twocolumn, 10pt]{IEEEtran}
\usepackage{stmaryrd}
\usepackage{amsfonts,amsmath,amssymb,amsthm}
\usepackage{times}
\usepackage{algorithm}
\usepackage{algorithmic}
\usepackage{bm}
\usepackage{balance}
\usepackage{color}
\usepackage{graphicx}
\usepackage{setspace}
\usepackage{comment}
\usepackage{cases}
\graphicspath{{./eps/}}
\usepackage{here}
\usepackage{array}
\usepackage{float}
\usepackage{cite}
\usepackage{url}
\usepackage{subfigure}

\newtheorem{myrem}{Remark}

\newcommand{\rfig}[1]{Fig.\,\ref{#1}} 
 
\newcommand{\req}[1]{\eqref{#1}} 
\newcommand{\bfmath}[1]{\mbox{\boldmath $#1$}} 
\newcommand{\rtab}[1]{Table\,\ref{#1}}

\newcommand{\rsec}[1]{Section\,\ref{#1}}

\newcommand{\ralg}[1]{Algorithm\,\ref{#1}}

\newcommand{\rline}[1]{line\,\ref{#1}}

\newcommand{\qedwhite}{\hfill \ensuremath{\Box}}

\begin{document}
\title{Learning self-triggered controllers with Gaussian processes}
\author{\large Kazumune~Hashimoto, Yuichi Yoshimura, Toshimitsu Ushio
\thanks{The authors are with the Graduate School of Engineering Science, Osaka University, Osaka, Japan (e-mail: kazumune.hashimoto@hopf.sys.es.osaka-u.ac.jp, yoshimura@hopf.sys.es.osaka-u.ac.jp, ushio@sys.es.osaka-u.ac.jp). The authors are supported by ERATO HASUO Metamathematics for Systems Design Project (No. JPMJER1603), JST.}
}
\maketitle

\begin{abstract}
This paper investigates the design of self-triggered controllers for networked control systems (NCSs), where the dynamics of the plant is \textit{unknown} apriori. 
To deal with the unknown transition dynamics, we employ the Gaussian process (GP) regression in order to learn the dynamics of the plant. To design the self-triggered controller, we formulate an optimal control problem, such that the optimal control and communication policies can be jointly designed based on the GP model of the plant. Moreover, we provide an overall implementation algorithm that jointly learns the dynamics of the plant and the self-triggered controller based on a reinforcement learning framework. Finally, a numerical simulation illustrates the effectiveness of the proposed approach.
\end{abstract}

\section{Introduction}
In networked control systems (NCSs), sensors, actuators, and controllers reside in multiple areas linked by wired/wireless communication network. Due to the progress in communication technology and many practical advantages such as a low-cost maintenance and flexibility for modifications, NCSs have been developed in a wide variety of applications, including manufacturing plants, autonomous robots/vehicles, traffic networks, to name a few\cite{ncssur2007a}. 
In recent years, event-triggered and self-triggered control have attracted much attention and are known to be useful strategies for the NCSs \cite{heemels2012a}. 
This is due to the fact that, it leads to the potential saving of resources that are present in NCSs, such as a limited battery capacity or a limited communication bandwidth, by transmitting sensor measurements over the communication network only when it is needed.  
So far, various event/self-triggered controllers have been proposed in the literature, see, e.g., \cite{event_survey} for survey papers. Early works consider designing event/self-triggered control based on input-to-state stability (ISS) or ${\cal L}_2$-gain performance\cite{lemmon2009a,heemels2011a,mazo2010}. 
More recently, event-triggered control has been formulated as the hybrid dynamical systems \cite{dolk2017c,heemels2013b}. 
In addition, some approaches to combine event/self-triggered control and optimal control have been also provided in recent years \cite{dimos2010a,hashimoto2017a,hashimoto2018c,hashimoto2017c,hashimoto2019b,hashimoto2019c,vam1,vam2,vam3,domagoj,liu,liu2}. 

In the aforecited event-triggered and self-triggered control framework, it is generally assumed that the transition dynamics, which represents the underlying model of the plant, is \textit{known} apriori. This implies that, when the event/self-triggered controllers are applied to the real world (actual) control systems, the resulting performance is heavily dependent on how the system model is accurate with respect to the true dynamics. However, it may be the case in practice when an accurate model of the plant is hard to obtain based on the first principles from physics, due to the fact that the dynamics is complex and highly nonlinear. 
Examples include mechanical systems \cite{modelfree4}, autonomous vehicles \cite{modelfree8}, power consumption of multi-story buildings \cite{modelfree3}, periodic errors in astrophotography systems \cite{modelfree7}, to name a few. 

Motivated by the above, in this paper we investigate the design of a novel self-triggered controller for NCSs, where the dynamics of the plant is assumed to be \textit{unknown} apriori. 
To this end, we make use of the Gaussian process (GP) regression \cite{rasmussen} in order to learn the dynamics of the plant. 
The use of GP offers many benefits, such as the ability to incorporate prior knowledge about the model (e.g., smoothness, periodicity) by selecting suitable kernel functions, as well as the ability to provide uncertainty of the model for prediction values. 
To design the self-triggered controller, we first formulate an infinite horizon optimal control problem, such that both the cost for the control performance and the communication are taken into account. Then, we derive the corresponding Bellman equation and provide an approach to solving the optimal control problem, such that both the optimal control and communication policies are designed based on the plant learned by the GP regression. {In particular, we employ a value iteration algorithm, which derives the optimal policies by iteratively improving the estimate of the optimal cost function. Moreover, when solving the value iteration algorithm, we employ the so-called \textit{moment matching} technique in order to approximate the multiple-ahead predictive distribution of states by the Gaussian distribution. As we will see later, this approximation together with the approximations of the optimal cost function based on the radial basis functions will allow us to derive the optimal policies in a tractable way.} 
Finally, we provide an overall implementation algorithm that jointly learns the dynamics of the plant as well as the optimal control and the communication policies based on a reinforcement learning framework. 
{As we will see later, this algorithm combines the \textit{exploration/exploitation phase} that aims at collecting the training data to learn the dynamics of the plant in an $\varepsilon$-greedy fashion, and the \textit{learning phase} that aims at updating the optimal control and communication policies based on the value iteration algorithm.} 

{In summary, the main contributions of this paper is provided as follows:
\begin{enumerate}
\item We formulate an infinite horizon optimal control problem, such that both the control and communication policies can be designed based on the GP model of the plant. 
\item 
We derive the Bellman equation corresponding to the optimal control problem and employ the value iteration algorithm to solve it.  
When solving this algorithm, we employ some approximation techniques, such as the moment matching, so that the (approximate) optimal policies can be derived. 
\item We provide an overall reinforcement learning algorithm that jointly learns the GP model of the plant as well as the optimal control and communication policies. 
\end{enumerate}}

\smallskip
{\textit{(Related works):}}
Our approach is related to several techniques that have been provided in the literature. Using the GP in control community has been attracted much attention in recent years \cite{modelfree4,modelfree8,modelfree3,modelfree7,modelfree5,modelfree6,modelfree10,bradford}. In particular, our approach is related to the ones based on optimal control framework, see, e.g., \cite{modelfree8,modelfree3,modelfree7,modelfree6,pilco,gpdp,modelfree10,bradford}. For example, in \cite{modelfree8}, the authors have utilized the GP model to learn the dynamics of the plant, and they have formulated a chance-constrained model predictive control (MPC), in which the optimal control problem is solved for each time step based on the knowledge about the dynamics learned by the GP. {In contrast to these previous methods, we provide an approach that jointly learns the dynamics of the plant and the \textit{self-triggered controller}, aiming at reducing the number of communication time steps for NCSs. As previously mentioned and will be clearer in later sections, this is achieved by formulating a value iteration algorithm, such that the optimal pair of the control input and the {inter-communication time steps} can be determined for each state based on the GP dynamics of the plant.}

With regard to the event/self-triggered control, some model-free/model-based approaches with unknown transition dynamics have been proposed in recent years, e.g.,  \cite{deepev,modelfree13,modelfree11,modelfree12,modelfree1,modelfree2,critic,vam4,vam5,vam6,vam7}.
 {For example, in \cite{modelfree1,modelfree2,critic,vam4,vam5,vam6}, an actor-critic based $Q$-learning algorithm was proposed to learn the intermittent feedback controller under the event-triggered policy, and closed-loop stability was rigorously shown.} {Our approach differs from those previous works, in the sense that; (i) we provide a model-based solution to the problem of learning self-triggered controllers based on the GP regression; (ii) while previous works aim at learning a controller based on a {prescribed structure} of the event-triggered condition (i.e., the event is triggered when the error between the actual state and the latest triggered state exceeds a certain threshold), our approach aims at learning both control and communication policies {from scratch}; (iii) while previous works deal with either linear or nonlinear input-affine systems, our approach is applicable to general nonlinear systems.}
{Moreover, in \cite{deepev}, a deep reinforcement learning was proposed to learn the event-triggered controller, and, similarly to our approach, the communication policy was designed from scratch. One of the potential advantages over this previous work may be that, since our approach is a model-based approach that explicitly incorporates the knowledge about the dynamics, it may require much fewer number of iterative tasks to learn the desired policies. 
Such data-efficiency (see, e.g., \cite{pilco}) is indeed illustrated in the simulation example in \rsec{simulation_sec}, where we show that the desired policies can be learned within $10$ episodes, while model-free approaches may typically require hundreds or thousands of iterative tasks to learn them.}

\smallskip
\smallskip 
\textit{\textbf{Notation.}} 
Throughout the paper, we make use of the following notations. 
Let $\mathbb{N}$, $\mathbb{N}_{\geq 0}$, $\mathbb{N}_{>0}$, $\mathbb{N}_{a:b}$ be the set of integers, non-negative integers, positive integers, and the set of integers in the interval $[a, b]$, respectively. 
Let $\mathbb{R}$, $\mathbb{R}_{\geq 0}$, $\mathbb{R}_{>0}$ be the set of reals, non-negative reals and positive reals, respectively. 
For a square matrix $\bfmath{Q}$, we use $\bfmath{Q} \succ 0$ to denote that $\bfmath{Q}$ is positive definite. Let ${\rm diag}(a_1,a_2, \ldots, a_N)$ be the diagonal matrix whose (diagonal) elements are given by $a_1, \ldots, a_N \in \mathbb{R}$. Moreover, let ${\rm Blkdiag}(A_1,A_2, \ldots, A_N)$ be the block diagonal matrix that consists of a set of matrices $A_1, \ldots, A_N$. 

\section{Preliminaries of Gaussian process regression}
In this section, we provide some basic concepts and useful properties of the Gaussian process (GP) regression. 
Consider a nonlinear function $h : \mathbb{R}^{n} \rightarrow \mathbb{R}$ expressed as
\begin{align}
y = h ({\bf x}) + \varepsilon, 
\end{align} 
where ${\bf x} \in \mathbb{R}^n$ is the input, $y \in \mathbb{R}$ is the output, and $\varepsilon \sim {\cal N} (0, \sigma^2 _{\varepsilon})$ is the Gaussian distributed white noise. 
In the GP regression, we assume that the function $h$ follows the GP. That is, for every set of a finite (or possibly infinite) number of inputs ${\bf x}_i \in \mathbb{R}^n$, $i=1, \ldots, N$, the joint probability of the corresponding set of outputs ${\bf y} = [y_1, y_2, \ldots, y_N]^\mathsf{T}$ follows the multivariate Gaussian distribution, i.e., ${\bf y} \sim \mathcal{N} (\mathbf{0}, \bfmath{K})$, where $\bfmath{K} \in \mathbb{R}^{N\times N}$ is the covariance matrix and is characterized by ${K}_{i j} = \mathsf{k} ({\bf x}_i, {\bf x}_j)$, where ${K}_{i j}$ is the $(i, j)$-component of $\bfmath{K}$ and $\mathsf{k} : \mathbb{R}^n \times \mathbb{R}^n \rightarrow \mathbb{R}_{\geq 0}$ is the positive definite kernel function.  

In this paper, we assume that the kernel function $\mathsf{k}$ is given by the squared exponential covariance function: 
\begin{equation}
\mathsf{k} ({\bf x}_i, {\bf x}_j) = \alpha^2 \exp \left (-\frac{1}{2} ({\bf x}_i - {\bf x}_j) ^\mathsf{T} \bfmath{\Lambda}^{-1} ({\bf x}_i - {\bf x}_j) \right ), 
\end{equation}
where $\bfmath{\Lambda} = {\rm diag} \left(\lambda^2 _1, \ldots, \lambda^2 _N\right)$ and $\{\alpha, \lambda_1, \ldots \lambda_N\}$ are the hyper-parameters. 
 For a given set of input-output training data ${\cal D} = \{{\bf x}_n, y_n\}^N _{n=1}$, the predictive distribution of the output for a new test input ${\bf x}$ follows the Gaussian distribution, i.e., $p(y | {\bf x}, {\cal D}) = {\cal N} (\mu ({\bf x}), \sigma ({\bf x}))$. Here the mean and the variance are given by  
\begin{align}
{\mu} ({\bf x} ) &= \bfmath{\mathsf{k}}^\mathsf{T} _* ({\bf x}) (\bfmath{K} + \sigma^2 _\varepsilon \bfmath{I})^{-1} {\bf y}, \\
{\sigma} ({\bf x} ) &= \mathsf{k}({\bf x} , {\bf x} ) - \bfmath{\mathsf{k}}^\mathsf{T} _* ({\bf x}) (\bfmath{K} + \sigma^2 _\varepsilon \bfmath{I})^{-1} \bfmath{\mathsf{k}}_* ({\bf x}), 
\end{align}
where ${\bf y} = [y_1, y_2, \ldots, y_N]^\mathsf{T}$ and 
\begin{align}
\bfmath{\mathsf{k}}_* ({\bf x}) = \left [\mathsf{k} ({\bf x},{\bf x}_1), \ldots, \mathsf{k} ({\bf x}, {\bf x}_N)\right]^\mathsf{T}.
\end{align} 
Suitable selections of the hyper-parameters $\{\alpha, \lambda_1, \ldots \lambda_N\}$ are given by evidence maximization, see, e.g., \cite{rasmussen}. 
For simplicity of presentation, we write $h \sim {\cal GP}$ if the function $h$ follows the GP.

\section{Problem statement}
{In this section, we describe the dynamics of the plant, overview of the self-triggered controller, and define the cost function to be minimized.} 
\subsection{Dynamics} \label{dynamics_sec}
\begin{figure}[t]
  \begin{center}
   \includegraphics[width=7.0cm]{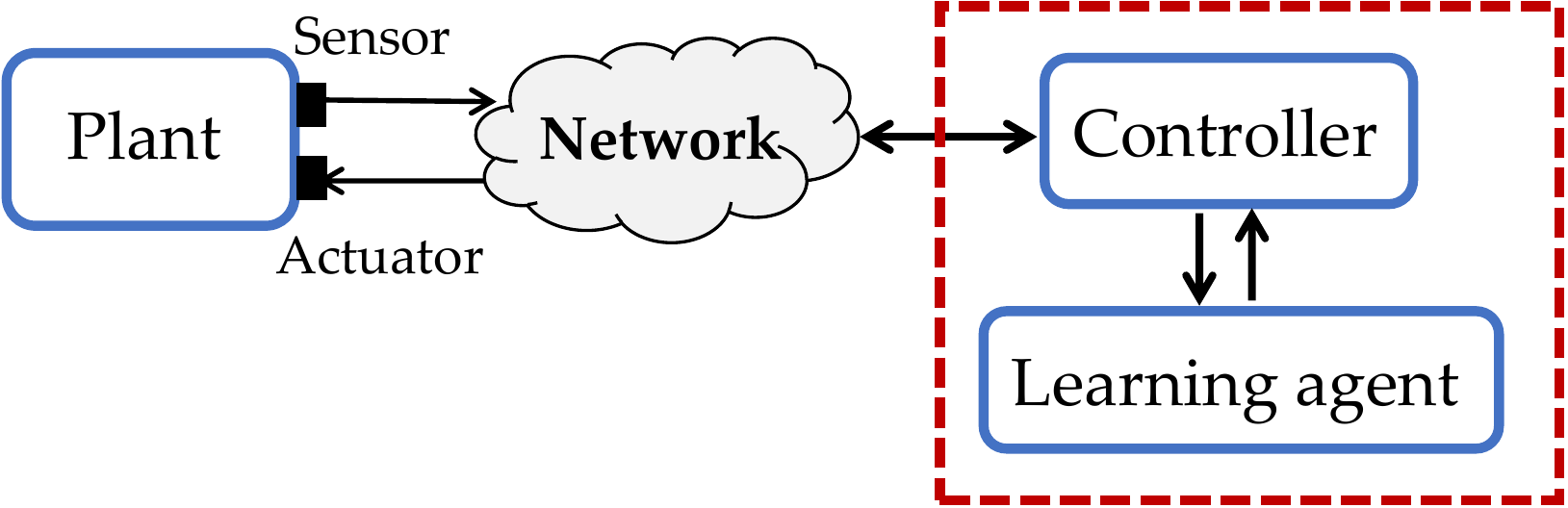}
   \caption{Networked Control System considered in this paper. } 
   \label{NCS}
  \end{center}
  \vspace{-0.5cm}
\end{figure}

We consider a networked control system (NCS) illustrated in \rfig{NCS}. 
As shown in the figure, the controller and the learning agent are connected to the plant over the communication network. 
Roughly speaking, the learning agent is responsible for learning the dynamics of the plant as well as the optimal control and communication policies. 
On the other hand, the controller is responsible for transmitting the control inputs to operate the plant based on the control and communication policies derived by the learning agent. 
This implementation will be formally given later in this paper. Throughout the paper, we assume that the communication network is ideal; it induces neither packet dropouts nor any network delays. 

The dynamics of the plant is given by the following nonlinear systems: 
\begin{equation}\label{dynamics}
{\bf x}_{k+1} = \bfmath{f}({\bf x}_k, {\bf u}_k), \ \ {\bf u}_k \in U,
\end{equation}
for all $k\in\mathbb{N}_{\geq 0}$, where ${\bf x}_k \in \mathbb{R}^{n_x}$ is the state, ${\bf u}_k \in \mathbb{R}^{n_u}$ is the control input, $U \subset \mathbb{R}^{n_u}$ is the set of control inputs, and $\bfmath{f} : \mathbb{R}^{n_x} \times \mathbb{R}^{n_u} \rightarrow \mathbb{R}^{n_x}$ is the transition dynamics that is assumed to be \textit{unknown} apriori. 
While the transition dynamics is unknown, it is assumed here that the equilibrium point is known; without loss of generality, we assume that the origin has the equilibrium point, i.e., $\bfmath{0} = \bfmath{f}(\bfmath{0}, \bfmath{0})$. The control goal is to stabilize the system towards the origin. 

Since $\bfmath{f}$ is unknown apriori, we consider that each component of the unknown function, i.e., $f_i$, $i\in \mathbb{N}_{1:n_x}$ ($\bfmath{f} = \left [f_1, f_2, \ldots, f_{n_x} \right]^\mathsf{T}$) is modeled by the GP regression. 
That is, $f_i$ is learned from the input-output training data ${\cal D}_i = \{\mathbf{X}, {\bf y}_i\}$, where 
\begin{align}
\mathbf{X} &= \left [
\left [
\begin{array}{cc}
{\bf x}^* _0 \\
{\bf u}^* _0
\end{array}
\right ], \left [
\begin{array}{cc}
{\bf x}^* _1 \\
{\bf u}^* _1
\end{array}
\right], \ldots, 
\left [
\begin{array}{cc}
{\bf x}^* _{N-1} \\
{\bf u}^* _{N-1}
\end{array}
\right ]
\right ], \label{training_data1} \\ 
{\bf y}_i &= [{x}^* _{i, 1}, {x}^* _{i, 2}, \ldots, {x}^* _{i, N} ]^\mathsf{T}.  \label{training_data2}
\end{align}

In \req{training_data1} and \req{training_data2}, 
$N \in \mathbb{N}_{>0}$ denotes the number of training data points, $[{{\bf x}^* _n}^\mathsf{T} , {{\bf u}^* _n} ^\mathsf{T}]$, $n \in \mathbb{N}_{1:N}$ are the training inputs following the dynamics \req{dynamics} 
(i.e., ${\bf x}^* _{n+1} = \bfmath{f} ({\bf x}^* _n, {\bf u}^* _n)$, $n \in \mathbb{N}_{0:N-1}$), and ${x}^* _{i, n}$, $i\in\mathbb{N}_{1:n_x}$ is the $i$-th element of ${\bf x}^* _n$ as the set of training outputs. We denote by $\mathsf{k}_i(\cdot, \cdot)$, $\bfmath{K}_i$ and $\{ \alpha_i, \lambda_{i, 1}, \ldots, \lambda_{i, N} \}$ the kernel function, covariance matrix and the hyper-parameters for the GP model of $f_i$, respectively. Moreover, we denote by ${\mu}_i ({{\bf x}}, {{\bf u}} )$, ${\sigma}_i ({{\bf x}}, {{\bf u}})$ the mean and the covariance for the GP model of $f_i$ with an arbitrary test input $\widetilde{{\bf x}} = [{{\bf x}}^\mathsf{T}, {{\bf u}}^\mathsf{T}]^\mathsf{T}$, respectively, i.e., 
\begin{align}
{\mu}_i ({{\bf x}}, {{\bf u}}) &= \bfmath{\mathsf{k}}^\mathsf{T} _{*,i} (\widetilde{\bf x}) (\bfmath{K}_i + \sigma^2 _\varepsilon \bfmath{I})^{-1} {\bf y}_i, \label{mui} \\
{\sigma}_i ({{\bf x}}, {{\bf u}}) &= \mathsf{k}_i (\widetilde{\bf x}, \widetilde{\bf x} ) - \bfmath{\mathsf{k}}^\mathsf{T} _{*, i} (\widetilde{\bf x}) (\bfmath{K}_i + \sigma^2 _\varepsilon \bfmath{I})^{-1} \bfmath{\mathsf{k}}_{*,i} (\widetilde{\bf x}), \label{sigmai}
\end{align}
where $\bfmath{\mathsf{k}}_{*,i} ({\bf x}) = \left [\mathsf{k}_i ({\bf x},{\bf x}_1), \ldots, \mathsf{k}_i ({\bf x}, {\bf x}_N)\right]^\mathsf{T}$. 
That is, letting $\widehat{f}_i$ be the GP model of $f_i$, we have 
\begin{align}\label{gpf}
\widehat{f}_i ({{\bf x}}, {{\bf u}}) \sim {\cal N} \left ({\mu}_i ({{\bf x}}, {{\bf u}}), {\sigma}_i ({{\bf x}}, {{\bf u}})\right).
\end{align} 
Then, the overall GP model for $\bfmath{f} = \left [f_1, f_2, \ldots, f_{n_x} \right]^\mathsf{T}$ is given by
\begin{align}\label{gpf2}
{\bfmath{\widehat{f}}} ({{\bf x}}, {{\bf u}}) \sim {\cal N} \left (\bfmath{\mu} ({{\bf x}}, {{\bf u}}), \bfmath{\Sigma} ({{\bf x}}, {{\bf u}})\right), 
\end{align}
where ${\bfmath{\widehat{f}}} = [\widehat{f}_1, \widehat{f}_2, \ldots, \widehat{f}_{n_x} ]^\mathsf{T}$ and 
\begin{align}
\bfmath{\mu} ({\bf x}, {\bf u}) &= \left[{\mu}_1 ({\bf x}, {\bf u}), \ldots, {\mu}_{n_x} ({\bf x}, {\bf u})\right]^\mathsf{T}, \\
\bfmath{\Sigma} ({\bf x}, {\bf u}) &= {\rm diag} \left({\sigma}_1 ({{\bf x}}, {{\bf u}} ), \ldots, {\sigma}_{n_x} ({{\bf x}}, {{\bf u}})\right).
\end{align}

\subsection{Overview of the self-triggered controller}
Let us now define the control and communication policies. First, let $k_{i}$, $i = 0, 1, 2, \ldots$ with $k_0 = 0$ and $k_{i+1} > k_i$, $\forall i\in\mathbb{N}_{\geq 0}$ be the communication time steps when the plant transmits the state $x_{k_i}$ to the controller. In addition, let $m_i \in \mathbb{N}_{>0}$, $i \in \mathbb{N}_{\geq 0}$ be the corresponding inter-communication time steps, i.e., $m_i = k_{i+1} - k_i$, $\forall i \in \mathbb{N}_{\geq 0}$. 
In this paper, we implement a \textit{self-triggered controller}\cite{heemels2012a}, aiming at reducing the number of communication time steps between the plant and the controller. That is, we aim at designing the {(deterministic) policies} $\pi = \{\pi_{\rm inp}, \pi_{\rm com}\}$, where 
\begin{itemize}
\item $\pi_{\rm inp} : \mathbb{R}^{n_x} \rightarrow \mathbb{R}^{n_u}$ is the \textit{control policy}, which is a mapping from the state to the corresponding control input; 
\item $\pi_{\rm com} : \mathbb{R}^{n_x} \rightarrow \mathbb{N}_{1:M}$ is the \textit{communication policy}, which is the mapping from the state to the corresponding inter-communication time steps. 
\end{itemize}
Here, $M \in \mathbb{N}_{>0}$ denotes the maximum inter-communication time step, which means that inter-communication time step does not exceed $M$. 
This parameter is a user-defined parameter and is chosen apriori in order to formulate the optimal control problem. The basic procedure of the self-triggered controller is summarized as follows: for each $k_i$, $i\in\mathbb{N}_{\geq 0}$, 
\begin{enumerate}
\renewcommand{\labelenumi}{[Step~\arabic{enumi}]}
\setlength{\leftskip}{0.5cm}
\item the plant measures the state ${\bf x}_{k_i}$ and transmits ${\bf x}_{k_i}$ to the controller; 
\item the controller computes the control input and the inter-communication time steps as ${\bf u} _{k_i} = \pi_{\rm inp} ({\bf x} _{k_i})$ and $m_i = \pi_{\rm com} ({\bf x} _{k_i})$;  
\item the controller transmits $\{ {\bf u} _{k_i}, m_i \}$ to the plant, and the plant applies ${\bf u} _{k_i}$ constantly until the next communication time, i.e., ${\bf u}_k = {\bf u} _{k_i}$, $\forall k\in\mathbb{N}_{k_i, k_{i+1}-1}$, where $k_{i+1} = k_i + m_i$; 
\end{enumerate}

\subsection{Cost function to be minimized}
In this paper, we consider the following infinite-horizon cost function to be minimized: 
\begin{equation}\label{cost}
J^\pi ({\bf x} _{k_i}) = \sum^\infty _{\ell = i+1}  \mathbb{E}^\pi _{{\bf x} _{k_\ell}} \bigl[ C_1 ({\bf x}_{{k_{\ell}}}) + \gamma C_2  (m_\ell) \bigr], 
\end{equation}
where $\mathbb{E}^\pi _{{\bf x}} [\cdot]$ denotes the expectation with respect to ${\bf x}$, $C_1 : \mathbb{R}^{n_x} \rightarrow \mathbb{R}_{\geq 0}$ represents the stage cost for the state, $C_2 : \mathbb{N}_{1:M} \rightarrow \mathbb{R}_{\geq 0}$ represents the {communication} cost that aims to penalize the inter-communication time steps, and  $\gamma > 0$ is the weight associated to the communication cost. 
We assume that the cost for the state is characterized by polynomials or exponential functions. For example, exponential type of the cost function is given by
\begin{align}
C_1({\bf x}_{{k_{\ell}}}) =  1 - \exp \left\{ -\frac{1}{2}{\bf x}_{{k_{\ell}}}^\mathsf{T} \bfmath{Q} {\bf x}_{{k_{\ell}}}\right\}, \label{state_cost}
\end{align}
where $\bfmath{Q} \succ 0$ is a given positive definite matrix. Moreover, polynomial cost functions include quadratic type: 
\begin{align}
C_1({\bf x}_{{k_{\ell}}}) = {\bf x}_{{k_{\ell}}}^\mathsf{T} \bfmath{Q} {\bf x}_{{k_{\ell}}}. \label{state_cost2}
\end{align}
As will be clearer in later sections, the above characterizations will allow us to provide analytical computations of the integrals with respect to the Gaussian probability distribution. 

The communication cost is characterized as follows: 
\begin{align}
C_2(m_\ell) = {M - m_\ell}. \label{com_cost}
\end{align}
Recall that $M$ is the maximum inter-communication time steps, i.e., $m_\ell \leq M, \forall \ell \in \mathbb{N}$. 
Hence, the total cost function defined in \req{cost} aims at taking the cost of the control performance and the communication into account, and the parameter $\gamma$ regulates the trade-off between them. As will be formalized in later sections, we design the optimal control and communication policies $\pi = \{\pi_{\rm inp}, \pi_{\rm com} \}$, such that \req{cost} is minimized. Note that, since the function $\bfmath{f}$ is unknown apriori and is learned by the GP regression, we will make use of the GP model $\bfmath{\widehat{f}}$ (see \req{gpf2}) in order to derive the optimal solution; for details, see \rsec{approximate_DP_sec}. 

\begin{myrem}[On the case of $\gamma = 0$]
\normalfont
{Note that, even for the case $\gamma = 0$, communication reduction can be potentially achieved by minimizing \req{cost}. This is due to the fact that the total cost in \req{cost} is defined by summing the stage costs \textit{only at the communication time steps}, i.e., the cost will be accumulated only when the communication is given. Hence, reducing the number of communication leads to the reduction of the total cost, and, therefore, minimizing \req{cost} leads to the communication reduction even for the case $\gamma = 0$. 
This interpretation will be also illustrated in the simulation example, where the communication reduction will be indeed achieved for the case $\gamma = 0$ in contrast to the time-triggered strategy; for details, see \rsec{simulation_sec}.} \qedwhite 
\end{myrem}

\section{Approximating Multiple-ahead predictions under constant control inputs}\label{approxmulti}
{In this section, we describe a way of how to approximate multiple-ahead predictions of states under constant control inputs, provided that the GP model of the plant is obtained.}  
Suppose that, for given GP model $\bfmath{\widehat{f}}$ in \req{gpf2} and
a pair $({\bf x}_k, {\bf u}) \in \mathbb{R}^{n_x} \times U$, 
we aim at computing the predictive distribution of the states with the constant control input ${\bf u}$, i.e., $p({\bf x}_{k+1} | {\bf x}_k, {\bf u}), p({\bf x}_{k+2} | {\bf x}_k, {\bf u}), \ldots$, where ${\bf x}_{k+m}$, $m \in \mathbb{N}_{>0}$ represent the state from ${\bf x}_k$ by applying ${\bf u}$ constantly for $m$ time steps. 
In this paper, we employ a moment matching technique \cite{pilco} in order to approximate the predictive distributions by the Gaussian distribution. 
Since the functions $f_i$, $i\in\mathbb{N}_{1:n_x}$ are modeled by the GP, the predictive distribution of the state for ${k+1}$ is given by $p({\bf x}_{k+1} | {\bf x}_k, {\bf u}) = {\cal N} (\bfmath{\mu}_{k+1}, \bfmath{\Sigma}_{k+1})$, where $\bfmath{\mu}_{k+1} = \bfmath{\mu} ({\bf x}_k, {\bf u}),\ \ \bfmath{\Sigma}_{k+1} = \bfmath{\Sigma}({\bf x}_k, {\bf u})$ with   
\begin{align}
\bfmath{\mu} ({\bf x}_k, {\bf u}) &= \left[{\mu}_1 ({\bf x}_k, {\bf u}), \ldots, {\mu}_{n_x} ({\bf x}_k, {\bf u})\right]^\mathsf{T} \\
\bfmath{\Sigma} ({\bf x}_k, {\bf u}) &= {\rm diag} \left({\sigma}_1 ({{\bf x}}_{k}, {{\bf u}} ), \ldots, {\sigma}_{n_x} ({{\bf x}}_{k}, {{\bf u}})\right).
\end{align} 
Here, ${\mu}_i (\cdot)$, ${\sigma}_i (\cdot)$ ($i \in \mathbb{N}_{1:n_x}$) are given by \req{mui} and \req{sigmai}, respectively. 
Now, suppose that we would like to compute the distribution of the predictive state for general $k + m$, $m=2,3, \ldots$. To this end, suppose that the predictive distribution of ${\bf x}_{k+\ell}$, $\ell \in \mathbb{N}_{1:m-1}$ is approximated by the Gaussian, i.e., $p({\bf x}_{k+\ell} | {\bf x}_k, {\bf u}) \approx {\cal N} (\bfmath{\mu}_{k+\ell}, \bfmath{\Sigma}_{k+\ell})$. 
Then, the predictive distribution for $k+\ell+1$ can be derived as follows: 
\begin{align}
p(&{\bf x}_{k+\ell+1} | {\bf x}_k, {\bf u}) \notag \\
                       & = \int p ( \widetilde{{\bf x}}_{k+\ell} | {\bf x}_k, {\bf u}) p ({\bf x}_{k+\ell+1} | \widetilde{{\bf x}}_{k+\ell}, {\bf x}_k, {\bf u}) {\rm d} \widetilde{{\bf x}}_{k+\ell}, \notag \\ 
                      & = \int p (\widetilde{{\bf x}}_{k+\ell} | {\bf x}_k, {\bf u}) p ({\bf x}_{k+\ell+1} | \widetilde{{\bf x}}_{k+\ell}) {\rm d} \widetilde{{\bf x}}_{k+\ell}, \label{twosteppred}
\end{align}
where we let $\widetilde{{\bf x}}_{k+\ell} = [{\bf x}^\mathsf{T} _{k+\ell}, {\bf u}^\mathsf{T} _{k+\ell}]^\mathsf{T}$ and ${\bf u}_{k+\ell}$ denotes the control input applied at $k+\ell$. 
Since the analytical computation of the integral in \req{twosteppred} cannot be given, we compute the mean and the covariance of the right hand side of \req{twosteppred} and approximate $p({\bf x}_{k+\ell+1} | {\bf x}_k, {\bf u})$ by the Gaussian distribution. 
The integral in \req{twosteppred} involves the joint distribution $p(\widetilde{{\bf x}}_{k+\ell} | {\bf x}_k, {\bf u})$, which is further computed as 
\begin{align}
p (\widetilde{{\bf x}}_{k+\ell} | {\bf x}_k, {\bf u}) &= p( {\bf x}_{k+\ell}, {\bf u}_{k+\ell} | {\bf x}_k, {\bf u}) \notag \\
& = p( {\bf x}_{k+\ell} | {\bf x}_k, {\bf u}) p({\bf u}_{k+\ell} | {\bf u}, {\bf x}_{k+\ell}) \notag 
\end{align}
Since ${\bf u}$ is applied constantly, it follows that ${\bf u}_{k+\ell} = {\bf u}$, i.e., $p({\bf u}_{k+\ell} | {\bf u}, {\bf x}_{k+\ell}) = p({\bf u}_{k+\ell} | {\bf u}) = {\rm Dirac} ({\bf u}_{k+\ell} - {\bf u})$, where ${\rm Dirac} (\cdot)$ denotes the Dirac delta function. 
Hence, \req{twosteppred} leads to 
\begin{align}
p&({\bf x}_{k+\ell+1} | {\bf x}_k, {\bf u}) \notag \\
  &= \int p( {\bf x}_{k+\ell} | {\bf x}_k, {\bf u}) p({\bf u}_{k+\ell} | {\bf u}) p ({\bf x}_{k+\ell+1} | \widetilde{{\bf x}}_{k+\ell}) {\rm d} \widetilde{{\bf x}}_{k+\ell} \notag \\
 &= \int p ({{\bf x}}_{k+\ell} | {\bf x}_k, {\bf u}) p ({\bf x}_{k+\ell+1} | {{\bf x}}_{k+\ell}, {\bf u}) {\rm d} {{\bf x}}_{k+\ell}, \label{twosteppred2} 
\end{align}
where $p ({{\bf x}}_{k+\ell} | {\bf x}_k, {\bf u}) \approx {\cal N} (\bfmath{\mu}_{k+\ell}, \bfmath{\Sigma}_{k+\ell})$. 
Moreover, using the GP model in \req{gpf2}, we have $p ({\bf x} _{k+\ell+1} | {{\bf x}}_{k+\ell}, {\bf u}) \approx p ({\bfmath{\widehat{f}}} ({{\bf x}}_{k+\ell}, {\bf u}) | {\bf x}_{k+\ell}, {\bf u}) = {\cal N} (\bfmath{\mu} ({\bf x}_{k+\ell}, {\bf u}), \bfmath{\Sigma} ({\bf x}_{k+\ell}, {\bf u}))$, where 
\begin{align}
\bfmath{\mu} &({\bf x}_{k+\ell}, {\bf u}) = \left[{\mu}_1 ({\bf x}_{k+\ell}, {\bf u}), \ldots, {\mu}_{n_x} ({\bf x}_{k+\ell}, {\bf u})\right]^\mathsf{T}, \notag  \\               
\bfmath{\Sigma} & ({\bf x}_{k+\ell}, {\bf u}) = {\rm diag} \left ({\sigma}_1 ({\bf x}_{k+\ell}, {\bf u}), \ldots, {\sigma}_{n_x} ({\bf x}_{k+\ell}, {\bf u}) \right ). \notag 
\end{align}
In the above, $\mu_i (\cdot)$ and $\sigma_i (\cdot)$ ($i \in \mathbb{N}_{1:n_x}$) are computed according to \req{mui} and \req{sigmai}, respectively. 
Based on the above, let us compute the mean and the covariance of the right hand side of \req{twosteppred2}. From \req{twosteppred2}, the mean of $p({\bf x}_{k+\ell+1} | {\bf x}_k, {\bf u})$ is given by 
\begin{align}
\bfmath{\mu}_{k+\ell+1} &= \mathbb{E}_{{{\bf x}}_{k+\ell}}\left[ \mathbb{E}_{{\bf x}_{k+\ell+1}}  [{\bf x}_{k+\ell+1}|{{\bf x}}_{k+\ell}, {\bf u}] \right] \notag \\
             & = \int p ({{\bf x}}_{k+\ell} | {\bf x}_k, {\bf u}) \bfmath{\mu} ({\bf x}_{k+\ell}, {\bf u}) {\rm d} {{\bf x}}_{k+\ell} \notag \\
             & = \int {\cal N}({\bfmath{\mu}}_{k+\ell}, {\bfmath{\Sigma}}_{k+\ell}) \bfmath{\mu} ({\bf x}_{k+\ell}, {\bf u}) {\rm d} {{\bf x}}_{k+\ell}. \label{mukm1}
\end{align}
The integral in \req{mukm1} can be computed analytically and is given by $\mu _{i, k+\ell+1} = \bfmath{\beta}_i^\mathsf{T} \bfmath{\eta}_i$, where $\mu _{i, k+\ell+1}$ denotes the $i$-th component of $\bfmath{\mu}_{k+\ell+1}$, $\bfmath{\beta}_i = (\bfmath{K}_i + \sigma^2 _\varepsilon \bfmath{I})^{-1} {\bf y}_i$ and 
$\bfmath{\eta}_i = [\eta_{i, 1}, \eta_{i, 2}, \ldots, \eta_{i, N}]$ with 
\begin{align}
\eta_{i, n} &= \alpha_i^2 \left|(\bfmath{\Lambda}_i)^{-1} \widetilde{\bfmath{\Sigma}}_{k+\ell} + \bfmath{I} \right|^{-1/2}\notag \\
&\times \exp \left (-\frac{1}{2} (\widetilde{\bfmath{\mu}}_{k+\ell} - \widetilde{\bf x}^* _n) ^\mathsf{T} (\bfmath{\Lambda}_i+\widetilde{\bfmath{\Sigma}}_{k+\ell})^{-1} (\widetilde{\bfmath{\mu}}_{k+\ell} - \widetilde{\bf x}^*_n) \right ),  \notag 
\end{align}
for all $i \in \mathbb{N}_{1:n_x}$, $n \in \mathbb{N}_{1:N}$. In the above, we let $\widetilde{\bfmath{\mu}}_{k+\ell} = [{\bfmath{\mu}}^\mathsf{T} _{k+\ell}, {\bf u}^\mathsf{T}]^\mathsf{T}$, $\widetilde{\bf x}^*_n =[{{\bf x}^* _n}^\mathsf{T}, {{\bf u}^* _n}^\mathsf{T}]^\mathsf{T}$ (recall that ${\bf x}^* _n$, ${{\bf u}^* _n}$ are the $n$-th training input defined in \req{training_data1}), and $\bfmath{\widetilde{\Sigma}}_{k+\ell} = {\rm Blkdiag} \left(\bfmath{{\Sigma}}_{k+\ell}, \bfmath{0}_{n_u\times n_u}\right)$ with $\bfmath{0}_{n_u\times n_u}$ being the $n_u\times n_u$ zero matrix. 
The covariance matrix $\bfmath{\Sigma}_{k+\ell+1}$ can be obtained by considering diagonal elements ${\sigma}_{i, k+\ell+1}$ and off-diagonal elements ${\sigma}_{ij, k+\ell+1}$, $i\neq j$ (see, e.g., \cite{pilco}). 
The diagonal elements are given by 
\begin{align}
{\sigma}_{i, k+\ell+1} & = \mathbb{E}_{{\bf x}_{k+\ell}}\left[{\rm Var}_{{\bf x}_{k+\ell+1}}[{x}_{i, k+\ell+1}|{\bf x}_{k+\ell}, {\bf u}]\right] \notag \\ 
&\ \ \ + {\rm Var}_{{\bf x}_{k+\ell}}\left[\mathbb{E}_{{\bf x}_{k+\ell+1}}[{x}_{i, k+\ell+1} | {\bf x}_{k+\ell}, {\bf u}]\right] \notag \\
& = \mathbb{E}_{{\bf x}_{k+\ell}}\left[{\rm Var}_{{\bf x}_{k+\ell+1}}[{x}_{i, k+\ell+1}|{\bf x}_{k+\ell}, {\bf u}]\right] \notag \\
&\ \ \ + \mathbb{E}_{{\bf x}_{k+\ell}}\left[\mathbb{E}^2_{{\bf x}_{k+\ell+1}}[{x}_{i, k+\ell+1} | {\bf x}_{k+\ell}, {\bf u}]\right]- \mu _{i, k+\ell+1} ^2 \notag \\
& = \bfmath{\beta}_i ^\mathsf{T} {\bfmath{L}}_i \bfmath{\beta}_i +  \alpha_i^2 - {\rm Tr} \left ( (\bfmath{K}_i + \sigma^2 _\varepsilon \bfmath{I} )^{-1} {\bfmath{L}}_i \right ) \notag \\ 
&\ \ \ +\sigma^2 _\varepsilon - \mu _{i, k+\ell+1} ^2, 
\end{align}
where ${\rm Var}_{\bf x}[\cdot]$ is the variance with respect to ${\bf x}$, ${\bfmath{L}}_i$ is the $N\times N$ matrix, whose $(p, q)$-component (denoted as $L_{i, pq}$) is given by 
\begin{align}
{{L}}_{i, pq} = & \left| \bfmath{R}_i \right|^{-1/2} \mathsf{k}_i (\widetilde{{\bf x}}^* _p, \widetilde{\bfmath{\mu}}_{k+\ell})\mathsf{k}_i (\widetilde{{\bf x}}^* _q, \widetilde{\bfmath{\mu}}_{k+\ell}) \notag \\ 
& \times \exp \left (2\bfmath{\Lambda}_i^{-2}  {({\widetilde{{\bf x}}_{p q}}^*)}^\mathsf{T} (\widetilde{\bfmath{\Sigma}}_{k+\ell}^{-1} + 2\bfmath{\Lambda}^{-1}_i )^{-1} \widetilde{{\bf x}}^* _{p q} \right ), \notag 
\end{align}
where $\widetilde{{\bf x}}^* _{p q} = \frac{1}{2}(\widetilde{{\bf x}}^* _p + \widetilde{{\bf x}}^* _q) - \widetilde{\bfmath{\mu}}_{k+\ell}$, $\bfmath{R}_i = 2 \bfmath{\Lambda}_i^{-1} \bfmath{\widetilde{\Sigma}}_{k+\ell} + \bfmath{I}$. 
The off-diagonal elements are given by 
\begin{align}
{\sigma}_{ij, k+\ell+1} = \bfmath{\beta}_i^\mathsf{T} {\bfmath{L}}_{ij} \bfmath{\beta}_j - \mu _{i, k+\ell+1} \mu _{j, k+m+1}, 
\end{align} 
where ${\bfmath{L}}_{ij}$ is the $N \times N$ matrix, whose $(p, q)$-component (denoted as ${{L}}_{ij, pq}$) is given by 
\begin{align}
&{{L}}_{ij, pq} = \left| \bfmath{R}_{ij} \right|^{-1/2} \mathsf{k}_i (\widetilde{{\bf x}}^* _p, \widetilde{\bfmath{\mu}}_{k+\ell})\mathsf{k}_j (\widetilde{{\bf x}}^* _q, \widetilde{\bfmath{\mu}}_{k+\ell}) \notag \\
&\times \exp \left (-\frac{1}{2}  \left(\widetilde{{\bf x}}^* _{pq, ij}\right)^\mathsf{T} \left(\bfmath{\Lambda}^{-1}_i + \bfmath{\Lambda}^{-1}_j + \widetilde{\bfmath{\Sigma}}_{k+\ell}^{-1}\right)^{-1} \widetilde{{\bf x}}^* _{pq, ij} \right ), \notag 
\end{align}
where $\bfmath{R}_{ij} = (\bfmath{\Lambda}_i ^{-1} + \bfmath{\Lambda}_j ^{-1})\bfmath{\widetilde{\Sigma}}_{k+\ell} + \bfmath{I}$ and 
\begin{align}
\widetilde{{\bf x}}^* _{pq, ij} = \bfmath{\Lambda}_i^{-1} (\widetilde{{\bf x}}^* _p - \widetilde{\bfmath{\mu}}_{k+\ell}) + \bfmath{\Lambda}_j^{-1} (\widetilde{{\bf x}}^* _q - \widetilde{\bfmath{\mu}}_{k+\ell}).
\end{align}
Based on the above, we can approximate $p({\bf x}_{k+\ell+1} | {\bf x}_k, {\bf u})$ by the Gaussian distribution as 
\begin{align}
p({\bf x}_{k+\ell+1} | {\bf x}_k, {\bf u}) \approx {\cal N} (\bfmath{\mu}_{k+\ell+1}, \bfmath{\Sigma}_{k+\ell+1}). 
\end{align}
Hence, by recursively applying the above procedure for all $\ell=1, \ldots, m-1$, we can approximate $p({\bf x}_{k+m} | {\bf x}_k, {\bf u})$ by the Gaussian distribution. 

\section{Approximate value iteration}\label{approximate_DP_sec}
{In this section, we provide an approach to deriving the optimal self-triggered controller that minimizes \req{cost}, provided the GP model of the plant \req{gpf2} is obtained.} 
Let $J^* ({\bf x}_{k_i}) = \min_{\pi} J^\pi ({\bf x}_{k_i})$. 
From \req{cost}, the corresponding optimal Bellman equation is given by 
\begin{align}\label{bellman}
&J^* ({\bf x}_{k_i}) \notag \\
&=  \min_{{\bf u}_{k_i}, m_{i}} \left \{ \mathbb{E}_{{\bf x}_{{k_{i+1}}}} \left [C({\bf x}_{{k_{i+1}}}, m_{i+1}) + J^* ({\bf x}_{{k_{i+1}}}) \right ] \right \}, 
\end{align}
where $C({\bf x}, m) = C_1 ({\bf x}) + \gamma C_2 (m)$. 
Since the state space $\mathbb{R}^{n_x}$ and the input space ${U}$ for the dynamics in \req{dynamics} are both infinite, deriving an explicit solution to \req{bellman} is in general intractable. 
Thus, 
we derive an approximated solution to \req{bellman} by employing a finite number of representative points in the state space and the input space, which are denoted as ${\bf x}_{R, 1}, {\bf x}_{R, 2}, \ldots, {\bf x}_{R, {N_X}} \in \mathbb{R}^{n_x}$ and ${\bf u} _{R, 1}, {\bf u} _{R, 2}, \ldots, {\bf u} _{R, {N_U}} \in U$, respectively, with $N_X$ and $N_U$ being the number of representative points. These representative points may be selected as the grid points in a given bounded region of $\mathbb{R}^{n_x}$ as well as $\mathbb{R}^{n_u}$, so that they include the origin (as we aim at stabilizing the state towards the origin). 
For simplicity of presentation, we let $\bfmath{X}_{R} = \left\{{\bf x}_{R, 0}, {\bf x}_{R, 1}, \ldots, {\bf x}_{R, {N_X}}\right\}$, $\bfmath{U}_{R} = \left\{{\bf u}_{R, 1}, {\bf u}_{R, 2}, \ldots, {\bf u}_{R, {N_U}}\right\}$. The optimal cost function (denoted as $\widehat{J}^*$) and the optimal control policy (denoted as $\widehat{\pi}^* _{\rm inp}$) are then approximated by the exponential Radial Basis Functions (RBFs): 
\begin{align}
\widehat{J}^* ({\bf x}) &= \sum^{N_{X}} _{n=1} w_{J, n} 
\exp \left(-\frac{1}{2 \sigma^2 _J} \left \| {\bf x} - {\bf x}_{R, n} \right \|^2 \right), \label{pi1}\\
\widehat{\pi}^* _{\rm inp} ({\bf x}) &= \sum^{N_{X}} _{n=1} w_{u, n} \exp \left(-\frac{1}{2 \sigma^2 _u} \left \| {\bf x} - {\bf x}_{R, n} \right \|^2 \right ), \label{pi2}
\end{align}
where $\{w_{J, n}\}^{N_X} _{n=1}$, $\{w_{u, n}\}^{N_X} _{n=1}$ are the weights and $\sigma_J$, $\sigma_u$ are the width of the RBFs for $\widehat{J}^*$ and $\widehat{\pi}^* _{\rm inp}$, respectively, which are the hyper-parameters to be designed and will be updated during the algorithm. Moreover, the optimal communication policy is approximated by $\widehat{\pi}^* _{\rm com} ({\bf x}) = \llbracket \widehat{\pi}' _{\rm com} ({\bf x}) \rrbracket$, where $\llbracket a \rrbracket$ denotes the closest positive integer to $a$ (i.e., $\llbracket a \rrbracket = \arg\min_j \{ |a - j| : j \in \mathbb{N}_{>0}\}$ ) and 
\begin{align}\label{pi3}
\widehat{\pi}' _{\rm com} ({\bf x}) = \sum^{N_X} _{n=1} w_{c, n} \exp \left(-\frac{1}{2 \sigma^2 _c} \left \| {\bf x} - {\bf x}_{R, n} \right \|^2 \right ). 
\end{align}
Here, $\{w_{c, n}\}^{N_X} _{n=1}$ and $\sigma _c$ are the hyper-parameters to be updated.

\begin{algorithm}[t]
 \caption{Approximate value iteration to derive the self-triggered controller.} \label{DP}
\begin{algorithmic}[1]
{\small
    \STATE Initialize the hyper-parameters to represent 
    $\widehat{\pi}^* _{\rm com}$, $\widehat{\pi}^* _{\rm inp}$ and $\widehat{J}^*$; 
    \FOR {${\rm Iteration}= 1: N_{{\rm ite}}$}
       \FOR {all ${\bf x} \in \bfmath{{X}}_{R}$} \label{algstart}
               \STATE $D^* ({\bf x} ) \leftarrow \infty$; 
                \FOR {all $({\bf u}, m) \in \bfmath{{U}}_{R} \times \mathbb{N}_{1:M}$}
                \STATE Compute $D ({\bf x}, {\bf u} , m)$ as follows:  \label{qevaluate}
           \begin{align*}
 D ({\bf x} , & {\bf u} , m) \notag \\
 & \longleftarrow \mathbb{E}_{{\bf x}_m} \left [C_1 ({\bf x} _m) +\gamma C_2(m') + \widehat{J}^* ({\bf x}_m) \right ]; 
           \end{align*} 
           \IF {$D ({\bf x} , {\bf u} , m) < D^* ({\bf x} )$}
            \STATE   $D^* ({\bf x} ) \leftarrow D ({\bf x}, {\bf u} , m)$; 
            \STATE    ${\bf u}^* ({\bf x} ) \leftarrow {\bf u}$;
            \STATE $m^* ({\bf x} ) \leftarrow m$;  \label{algend}
            \ENDIF
            \ENDFOR
            \ENDFOR
        \STATE Update the hyper-parameters to represent $\widehat{\pi}' _{\rm com}$, $\widehat{\pi}^* _{\rm inp}$ and $\widehat{J}^*$ by using the new training data: 
        \begin{align}
        &\left\{ {\bf x}_{R, n}, {D}^* ({\bf x}_{R, n})\right\}^{N_X} _{n=1}, \left\{ {\bf x}_{R, n}, {\bf u}^* ({\bf x}_{R, n})\right \}^{N_X} _{n=1}, \notag \\ 
&\left \{ {\bf x}_{R, n}, {m}^* ({\bf x}_{R, n})\right\}^{N_X} _{n=1};
        \end{align}
        \ENDFOR
        \smallskip}
\end{algorithmic}
\end{algorithm}

The iterative procedure to solve \req{bellman} follows the so-called \textit{value iteration}\cite{bertsekas}, which is summarized in \ralg{DP}. 
As shown in the algorithm, for each ${\bf x}  \in \bfmath{{X}}_{R}$, we compute $D( {\bf x} , {\bf u} , m)$ for all ${\bf u} \in \bfmath{{U}}_{R}$ and $m\in\mathbb{N}_{1:M}$, which are specifically defined as 
\begin{align}
 D ({\bf x}, {\bf u} , m) &= \mathbb{E}_{{\bf x} _m} \left [C_1 ({\bf x} _m) + \gamma C_2 (m')  + \widehat{J}^* ({\bf x} _m) \right ] \notag \\
               & = \int p({\bf x}_m | {\bf x}, {\bf u}) C_1 ({\bf x} _m) {\rm d} {\bf x}_m \label{first} \\
               &\ \ \ \  + \gamma \int p({\bf x} _m | {\bf x}, {\bf u}) C_2 (m') {\rm d} {\bf x} _m \label{second}  \\
              &\ \ \ \ + \int p({\bf x} _m | {\bf x}, {\bf u}) \widehat{J}^* ({\bf x} _m) {\rm d} {\bf x} _m \label{third}
\end{align}
where ${\bf x}_m$ is the state that is reached from ${\bf x}$ by applying ${\bf u}$ constantly for $m$ time steps, $m'$ is the inter-communication time steps determined for the state ${\bf x}_m$, i.e., $m' = \widehat{\pi}^* _{\rm com} ({\bf x} _m)$. 
As shown in \req{first}--\req{third}, it is required to compute the distribution $p({\bf x} _m | {\bf x}, {\bf u})$, as well as the three expected values (integrals) with respect to this distribution. In what follows, we provide a detailed way of computing these three terms.  

\smallskip
\textit{(Computation of  $p({\bf x} _m | {\bf x}, {\bf u})$):} The term $p({\bf x}_m | {\bf x}, {\bf u})$ is the predictive distribution of the state from ${\bf x}$ by applying ${\bf u}$ constantly for $m$ time steps, which can be indeed approximated by the moment matching technique as discussed in \rsec{approxmulti}. 
That is, we can approximate the distribution as $p({\bf x} _m | {\bf x}, {\bf u}) \approx {\cal N} (\bfmath{\mu}_{m}, \bfmath{\Sigma}_{m})$, where $\bfmath{\mu}_{m}$ and $\bfmath{\Sigma}_{m}$ denote the mean and the covariance of $p({\bf x}_m | {\bf x}, {\bf u})$ that are computed by following the technique described in \rsec{approxmulti}. 

\smallskip
\textit{(Computation of \req{first}):} Using the Gaussian approximation of $p({\bf x}_m | {\bf x}, {\bf u})$, the first term \req{first} is given by
\begin{align}
\int p({\bf x}_m &| {\bf x}, {\bf u}) C_1 ({\bf x}_m) {\rm d}{\bf x}_m  \notag \\
&\approx \int {\cal N} (\bfmath{\mu}_{m}, \bfmath{\Sigma}_{m}) C_1({\bf x}_m) {\rm d}{\bf x}_m. \label{first_integral} 
\end{align}
Since we assume that $C_1$ is characterized by polynomials or exponential, we can analytically compute the integral in \req{first_integral}. 
For example, if $C_1$ is given by \req{state_cost}, the integral in \req{first_integral} further leads to 
\begin{align}
\int& {\cal N} (\bfmath{\mu}_{m}, \bfmath{\Sigma}_{m}) C_1 ({\bf x}_m) {\rm d}{\bf x}_m \notag \\
 & = 1-\int {\cal N} (\bfmath{\mu}_{m}, \bfmath{\Sigma}_{m}) \exp \left\{ -\frac{1}{2}{\bf x}_m^\mathsf{T} \bfmath{Q} {\bf x}_m \right\} {\rm d}{\bf x}_m \notag \\ 
 & = 1 -  \delta  (\bfmath{\mu}_{m},\bfmath{\Sigma}_m) \notag 
\end{align}
where $\delta (\cdot)$ is given by
$\delta(\bfmath{\mu}_{m}, \bfmath{\Sigma}_m) = |\bfmath{I} + \bfmath{\Sigma}_m \bfmath{Q}|^{-\frac{1}{2}}$ $\exp \left (- \frac{1}{2} \bfmath{\mu}_{m} \bfmath{Q} (\bfmath{I} + \bfmath{\Sigma}_{m} \bfmath{Q})^{-1} \bfmath{\mu}_{m} \right )$. 

\smallskip
\textit{(Computation of \req{second})}: The second term \req{second} can be computed as 
\begin{align}
\int p({\bf x}_m | & {\bf x}, {\bf u}) C_2 (m') {\rm d} {\bf x}_m \notag \\ 
              & = \int p({\bf x}_m | {\bf x}, {\bf u}) ({M - \widehat{\pi}^* _{\rm com} ({\bf x}_m)}) {\rm d} {\bf x}_m \notag \\ 
              & = M - \int p({\bf x}_m | {\bf x}, {\bf u}) \widehat{\pi}^* _{\rm com} ({\bf x}_m) {\rm d} {\bf x}_m. 
\end{align}
which requires to compute $\int p({\bf x}_m | {\bf x}, {\bf u}) \widehat{\pi}^* _{\rm com} ({\bf x}_m) {\rm d} {\bf x}_m$. 
Using \req{pi3}, we approximate this term as follows: 
\begin{align}
\int p( &{\bf x}_m | {\bf x}, {\bf u}) \widehat{\pi}^* _{\rm com} ({\bf x}_m) {\rm d} {\bf x}_m \notag \\ 
&\ \ \approx \int p({\bf x}_m | {\bf x}, {\bf u}) \widehat{\pi}' _{\rm com} ({\bf x}_m) {\rm d} {\bf x}_m \notag \\ 
&\ \  = \sum^{N_{X}} _{n=1} w_{c, n} \delta_{c, n} (\bfmath{\mu}_m, \bfmath{\Sigma}_m) \label{delta}
\end{align}
where $\delta_{c, s}(\cdot)$ is given by $\delta_{c,s} (\bfmath{\mu}_m, \bfmath{\Sigma}_m) = |\bfmath{I} + \sigma^{-2} _c \bfmath{\Sigma}_m|^{-\frac{1}{2}} $ $\exp \left (- \frac{1}{2\sigma^2 _c} (\bfmath{\mu}_m - {\bf x}_{R, n} )^\mathsf{T} (\bfmath{I} + \bfmath{\Sigma}_m \sigma^{-2} _c)^{-1} (\bfmath{\mu}_m - {\bf x}_{R, n} ) \right )$. 

\smallskip 
\smallskip
\smallskip
\textit{(Computation of \req{third}):} The third integral \req{third} can be approximated in a similar manner to the computation of \req{second}. 
From \req{pi1} and using $p({\bf x} _m | {\bf x}, {\bf u}) \approx {\cal N} (\bfmath{\mu}_{m}, \bfmath{\Sigma}_{m})$, we have 
\begin{align}
\int p( {\bf x}_m | {\bf x}, {\bf u}) \widehat{J}^* ({\bf x}_m) {\rm d} {\bf x}_m = \sum^{N_{X}} _{n=1} w_{J, n} \delta_{J, n} (\bfmath{\mu}_m, \bfmath{\Sigma}_m), 
\end{align}
where $\delta_{J, n}(\cdot)$ is given by $\delta_{J, n} (\bfmath{\mu}_m, \bfmath{\Sigma}_m) = |\bfmath{I} + \sigma^{-2} _J \bfmath{\Sigma}_m| $ $\exp \left (- \frac{1}{2\sigma^2 _J} (\bfmath{\mu}_m - {\bf x}_{R, n} )^\mathsf{T} (\bfmath{I} + \bfmath{\Sigma}_m\sigma^{-2} _J)^{-1} (\bfmath{\mu}_m - {\bf x}_{R, n} ) \right )$. 

\smallskip
\smallskip
As shown in the algorithm (\rline{algstart}--\rline{algend}), for each ${\bf x} \in \bfmath{{X}}_{R}$ 
we pick the smallest value among $D({\bf x}, {\bf u}, m)$, ${\bf u} \in \bfmath{{U}}_{R}$, $m\in\mathbb{N}_{1:M}$, as well as the corresponding pair of the control input and inter-communication time steps, which we denote by $D^* ({\bf x})$, ${\bf u}^* ({\bf x})$, and $m^* ({\bf x})$, respectively. Consequently, we obtain 
$\left \{{D}^* ({\bf x}_{R, n}), {\bf u}^* ({\bf x}_{R, n}), m^* ({\bf x}_{R, n}) \right \}^{N_X} _{n=1}$, and these are used as the new training data to update the hyper-parameters of $\widehat{J}^*$, $\widehat{\pi}^* _{\rm inp}$, $\widehat{\pi}' _{\rm com}$ in \req{pi1}, \req{pi2}, \req{pi3}. For example, $\widehat{J}^*$ is updated by using the training data $\left\{ {\bf x}_{R, n}, {D}^* ({\bf x}_{R, n})\right\}^{N_X} _{n=1}$, where ${\bf x}_{R, n}$, $n\in\mathbb{N}_{1:N_X}$ are the training inputs and ${D}^* ({\bf x}_{R, n})$,  $n\in\mathbb{N}_{1:N_X}$ are the training outputs. 

\begin{myrem}[On the selection of $M$]
\normalfont
{As $M$ is selected larger, we may achieve a more communication reduction, since the controller may increase the possibility to select larger inter-communication time steps. 
However, the execution time to derive the optimal policies may increase as $M$ is selected larger, due to the fact that the number of evaluations to compute $D ({\bf x} , {\bf u} , m)$ (\rline{qevaluate} in \ralg{DP}) increases. 
Hence, the parameter $M$ may be carefully chosen by considering the tradeoff between the communication reduction for the NCS and the computation load to derive the optimal policies according to \ralg{DP}. \qedwhite }
\end{myrem}


\section{Implementation} 
In this section, we provide an overall implementation algorithm that jointly learns the dynamics of the plant and the self-triggered controller based on a reinforcement learning framework. 

\begin{figure}[t]
  \begin{center}
   \includegraphics[width=8.6cm]{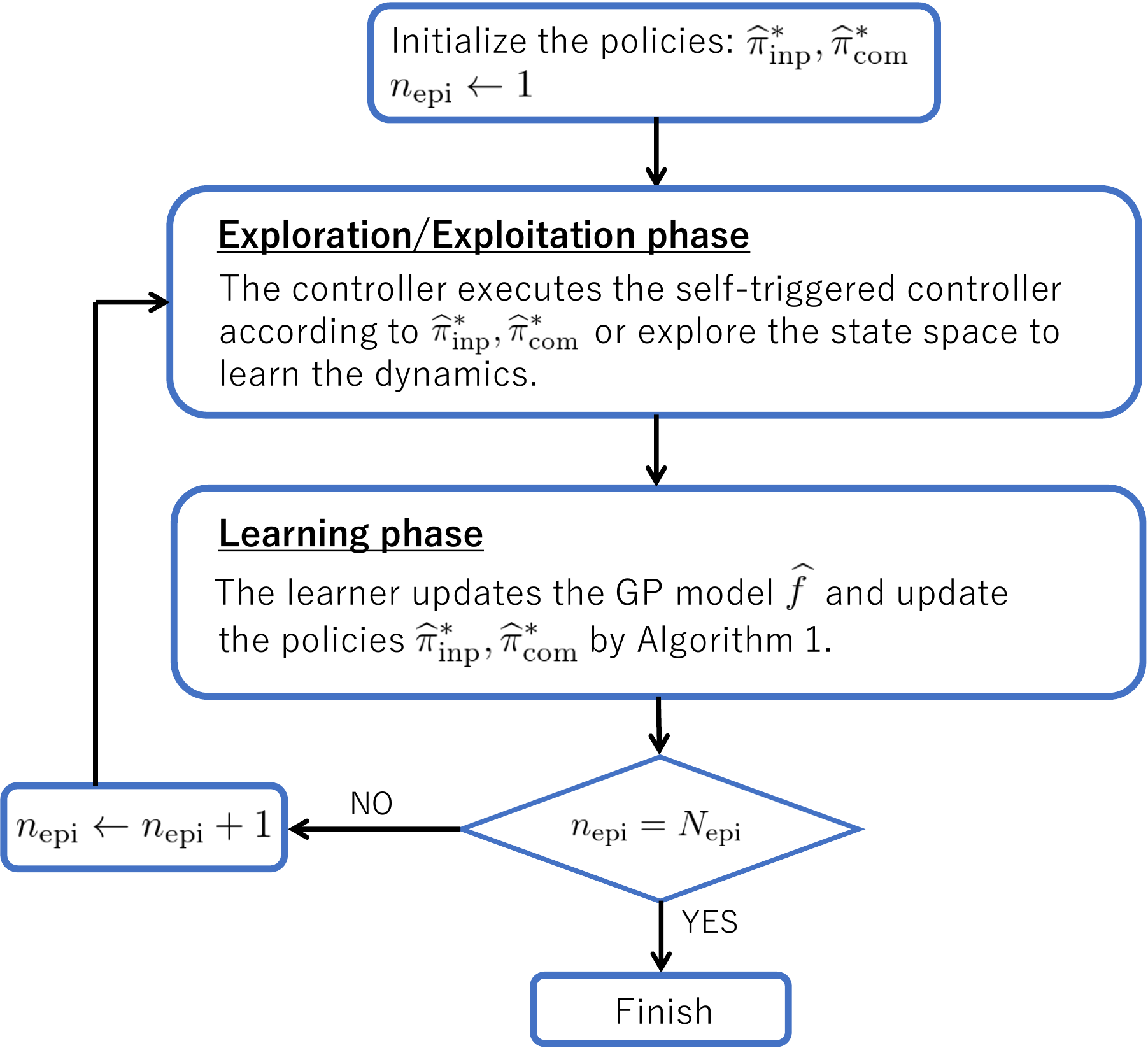}
   \caption{{Flowchart of the overall algorithm (\ralg{online}). As shown in the figure, the algorithm mainly consists of the \textit{exploration/exploitation phase} that aims at executing the self-triggered controller or collecting the training data to learn the dynamics of the plant, and the \textit{learning phase} that aims at updating the GP model $\bfmath{\widehat{f}}$ as well as the optimal control and communication policies based on the value iteration algorithm (\ralg{DP}).}} 
   \label{flowfig}
  \end{center}
  \vspace{-0.4cm}
\end{figure}

\renewcommand{\algorithmicrequire}{\textbf{Input:}}
\renewcommand{\algorithmicensure}{\textbf{Output:}}
\begin{algorithm}[t]
\caption{Overall reinforcement learning algorithm.}\label{online}
\begin{algorithmic}[1]
{\small 
\REQUIRE{${\bf x}_{\rm init}$ (initial state), $N_{\rm epi}$ (number of episodes), $\varepsilon \in [0, 1)$ (threshold for the greedy policy);}
\ENSURE{$\widehat{\pi}^* _{\rm inp}$, $\widehat{\pi}^* _{\rm com}$ (approximated optimal control and communication policies);}
\smallskip
 \STATE   Initialize the hyper-parameters to represent 
    $\widehat{\pi}^{*} _{\rm inp}$, and set $\widehat{\pi}^{*} _{\rm com} ({\bf x}) \leftarrow 1$, $\forall {\bf x} \in \mathbb{R}^{n_x}$; \label{initialalg}
  \STATE  $\mathbf{X} \leftarrow \{ \}$, $\forall i \in \mathbb{N}_{1:n_x}$; 
  \STATE ${\bf y}_i \leftarrow \{ \}$, $\forall i \in \mathbb{N}_{1:n_x}$; 
 \STATE  ${\cal D}_i \leftarrow \left\{\mathbf{X}, {\bf y}_i\right\}$,  $\forall i \in \mathbb{N}_{1:n_x}$ (initialize the training data); \smallskip
    \FOR{$n_{\rm epi} = 1: N_{\rm epi}$} 
  \STATE  $\ell \leftarrow 0$; 
  \STATE $k_\ell \leftarrow 0$; 
   \STATE ${\bf x}_{k_\ell} = {\bf x}_{\rm init}$; 
  \STATE  The plant transmits ${\bf x}_{k_\ell}$ to the controller; \smallskip 
  \STATE \underline{{[{Exploration/Exploitation phase}]}}  \smallskip
       \FOR {$\ell = 0: N_{\max}-1$} \label{algstart2}
        \STATE Sample $r \sim {\rm Uniform} [0, 1]$; \label{uniformline} 
       \IF {$r < \varepsilon$}
            \STATE ${m}_{\ell} \leftarrow 1$; 
            \STATE Select ${\bf u}_{k_\ell}$ randomly from $U$; 
               \ELSE
               \STATE ${\bf u}_{k_\ell} \leftarrow \widehat{\pi}^* _{\rm inp} ({\bf x}_{k_\ell})$; 
               \STATE ${m}_{\ell} \leftarrow \widehat{\pi}^* _{\rm com} ({\bf x}_{k_\ell})$;
               \ENDIF
               \ENDFOR
              \STATE $k_{\ell+1} \leftarrow k_\ell + m_\ell$;
              \STATE The controller transmits $\{{\bf u}_{k_\ell}, m_\ell\}$ to the plant; 
              \STATE The plant applies ${\bf u}_{k_\ell}$ constantly for $m_\ell$ time steps and transmit ${\bf x}_{k_{\ell+1}}$ to the controller; 
               \IF{ $m_\ell = 1$ } 
            \STATE   $\mathbf{X} \leftarrow \{\mathbf{X} \cup [{\bf x}^\mathsf{T} _{k_\ell}, {\bf u}^\mathsf{T} _{k_\ell}]^\mathsf{T}\}$;
            \STATE    ${\bf y}_i \leftarrow \{ {\bf y}_i \cup {x}_{k_{\ell+1}, i} \}$, $\forall i \in \mathbb{N}_{1:n_x}$;
             \STATE ${\cal D}_i \leftarrow \{{\cal D}_i \cup \{\mathbf{X}, {\bf y}_i\}\}$;
               \ENDIF
               \label{algend2} \smallskip     
     \STATE  \underline{{[Learning phase]}}  \smallskip
     \STATE  The learning agent learns the GP model of the plant by using the new training data ${\cal D} = \left \{{\cal D}_i\right\}^{n_x} _{i=1}$; \label{updatagpalg}
      \STATE The learning agent executes \ralg{DP} to update the (approximated) optimal policies $\widehat{\pi}^* _{\rm inp}$, $\widehat{\pi}^* _{\rm com}$; \label{updatagpalg2}
    \ENDFOR}
\end{algorithmic} 
\end{algorithm}

{The overall algorithm is shown in \rfig{flowfig} as a flowchart and the details are shown in \ralg{online}.} 
Since we assume that the learning agent has no knowledge about the dynamics of the plant, we set the communication policy as $\widehat{\pi}^* _{\rm com} ({\bf x}) \leftarrow 1$, $\forall {\bf x} \in \mathbb{R}^{n_x}$ (i.e., communication is given at every time step), so that the learning agent can efficiently collect the training data and learn the dynamics of the plant at the initial phase (\rline{initialalg} in \ralg{online}). 
As shown in \rfig{flowfig} and \ralg{online}, the algorithm mainly consists of the following two steps; exploration/exploitation phase (\rline{algstart2}--\rline{algend2} in \ralg{online}), and learning phase (\rline{updatagpalg}, \rline{updatagpalg2} in \ralg{online}). 
During the exploration/exploitation phase, 
the controller implements the self-triggered controller in an $\varepsilon$-greedy fashion, as well as updates the training data. 
In the algorithm, ${\rm Uniform} (0, 1)$ (\rline{uniformline}) generates a random real number from the interval $[0, 1]$ according to the uniform distribution. 
That is, with the probability $\varepsilon$, a random control input with the one step inter-communication time step is sampled, and, otherwise, the computed optimal control and communication policies are chosen to be executed. Here, the one step inter-communication time step is chosen (with the probability $\varepsilon$) so that the learning agent is able to utilize the consecutive states (i.e., ${\bf x}_{k_\ell}$, ${\bf x}_{k_\ell+1}$ with $k_{\ell+1} = k_\ell + 1$) to update the GP model of $\bfmath{f}$. In the learning phase, the learning agent utilizes the new training data ${\cal D}$ to update the GP model of the plant, and compute the (approximated) optimal control and communication policies according to \ralg{DP}. 

Finally, some remarks on the proposed algorithm are in order as follows:
\begin{myrem}[On achieving closed-loop stability]
\normalfont
{
Proving closed-loop stability by the proposed approach (\ralg{online}) is indeed challenging due to the following reasons. First, since we include the cost of communication in \req{cost}, if $\gamma$ (i.e., the weight for the communication cost) is selected too large, the penalty of the communication is too emphasized and the convergence to the origin may not be guaranteed. Indeed, this issue will be pointed out in the simulation result, where it is shown that, as $\gamma$ is selected larger, the state does not converge to origin (for details, see Section~VII). }
{The closed-loop stability may be achieved as $\gamma \rightarrow 0$. 
However, since we employ the GP model of the plant when solving the optimal control problem, we first need to show that the GP model of the plant is accurate enough with respect to the true (actual) dynamics. 
Since there is no theoretical result on the error bound between the GP model $\widehat{f}$ and the true one $f$, how much training data should be collected to obtain the accurate model may be in general unknown. Hence, even though there exists a self-triggered controller that stabilizes the actual system to the origin, such stabilization is not guaranteed according to the policies derived according to \ralg{online}.} \qedwhite 
\end{myrem}
\begin{myrem}
\normalfont
{The lack of providing theoretical proof on closed-loop stability may be the drawback of our approach with respect to some previous works of event-triggered control with unknown transition dynamics (see, e.g., \cite{modelfree1,modelfree2,critic,vam4,vam5,vam6}). 
Nevertheless, our approach is advantageous over these previous works, in the sense that our approach is applicable to \textit{general} nonlinear systems, while previous works focus on only input-affine or linear systems. 
For example, the prescribed event-triggered condition may be difficult to characterize for general nonlinear systems based on the procedure presented in \cite{critic}, due to the fact that the Hamilton-Jacobi-Bellman (HJB) equation under the event-triggered strategy is no longer characterized by (13) in \cite{critic}. In this paper, the self-triggered controller for general nonlinear systems can be designed by learning the dynamics based on the GP regression and deriving both the control and communication policies from scratch by implementing \ralg{DP}.}\qedwhite 
\end{myrem}
\section{Simulation results} \label{simulation_sec}
In this section, we illustrate the effectiveness of the proposed approach through a simulation example. 
The simulation was conducted on Matlab 2016a under Windows 10, Intel(R) Core(TM) i7 4.20 GHz, 32 GB RAM. As a simulation example, we consider a control problem of an inverted pendulum, whose dynamics is governed by 
\begin{align}
{x}_{1,k+1} &=  x_{1,k} + \Delta x _{2,k} \label{ex1}\\
{x}_{2, k+1} & = x_{2,k} + \Delta ( \sin x_{1,k} - x_{2,k} + u_k), \label{ex2}
\end{align}
where $x_{1, k}$ and $x_{2, k}$ with ${\bf x}_{k} = [x_{1, k}; x_{2, k}]$ are the states that represent the angular position and the velocity of the mass, $u_k \in \mathbb{R}$ is the control input, and $\Delta = 0.2$ denotes the sampling time interval. 
{Letting $\bfmath{f}({\bf x}_k, {u}_k) = [f_1({\bf x}_k, {u}_k), f_2({\bf x}_k, {u}_k)]^\mathsf{T}$ with $f_1({\bf x}_k, {u}_k) = x_{1, k} + \Delta x_{2,k}$ and $f_2({\bf x}_k, {u}_k) = x_{2,k} + \Delta ( \sin x_{1,k} - x_{2,k} + u_k)$, we obtain the discrete-time system as ${\bf x}_{k+1} = \bfmath{f}({\bf x}_k, {u}_k)$, $k\in\mathbb{N}$. Note that the function $\bfmath{f}(\cdot, \cdot)$ is assumed to be unknown apriori and is thus learned by the GP regression.}
It is assumed that $U = [-1.5, 1.5]$ and the initial state is given by ${\bf x}_{\rm init} = [{x}_{1, 0}; {x}_{2, 0}] = [1.0; 0.2]$. 
The maximum inter-communication time step is $M=10$, and the representative points for the state space to solve \req{bellman} is selected by the uniform grid points in the set ${X} = [-1.5, 1.5]\times [-1.5, 1.5]$ with the interval $0.3$, i.e., ${X}_{R} = [{X}]_{0.3}$. The representative points for the input space is given by ${U}_{R} = [{U}]_{0.3}$. We use the exponential type for the stage cost in \req{state_cost} with $\bfmath{Q} = \bfmath{I}_2$, and we set $\gamma = 0$ for the cost function in \req{cost}. 

\begin{figure}[t]
   \centering
    \subfigure[State trajectories by implementing \ralg{online} with $M=10$.]{
      {\includegraphics[width=8.5cm]{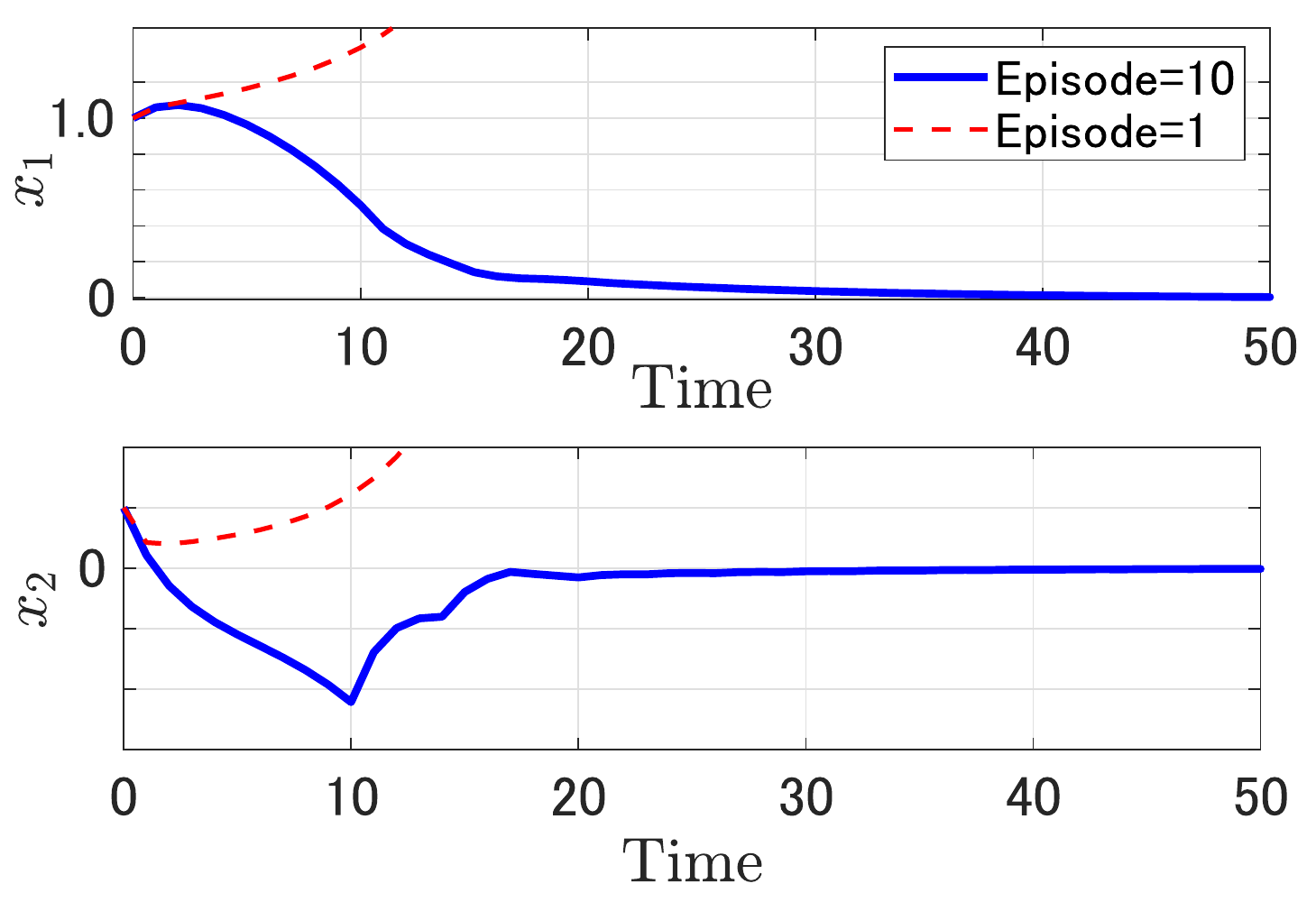}} \label{trajectoryfig}}
    \subfigure[Corresponding inter-communication time steps (${\rm Episode} = 10$).]{
      {\includegraphics[width=8.5cm]{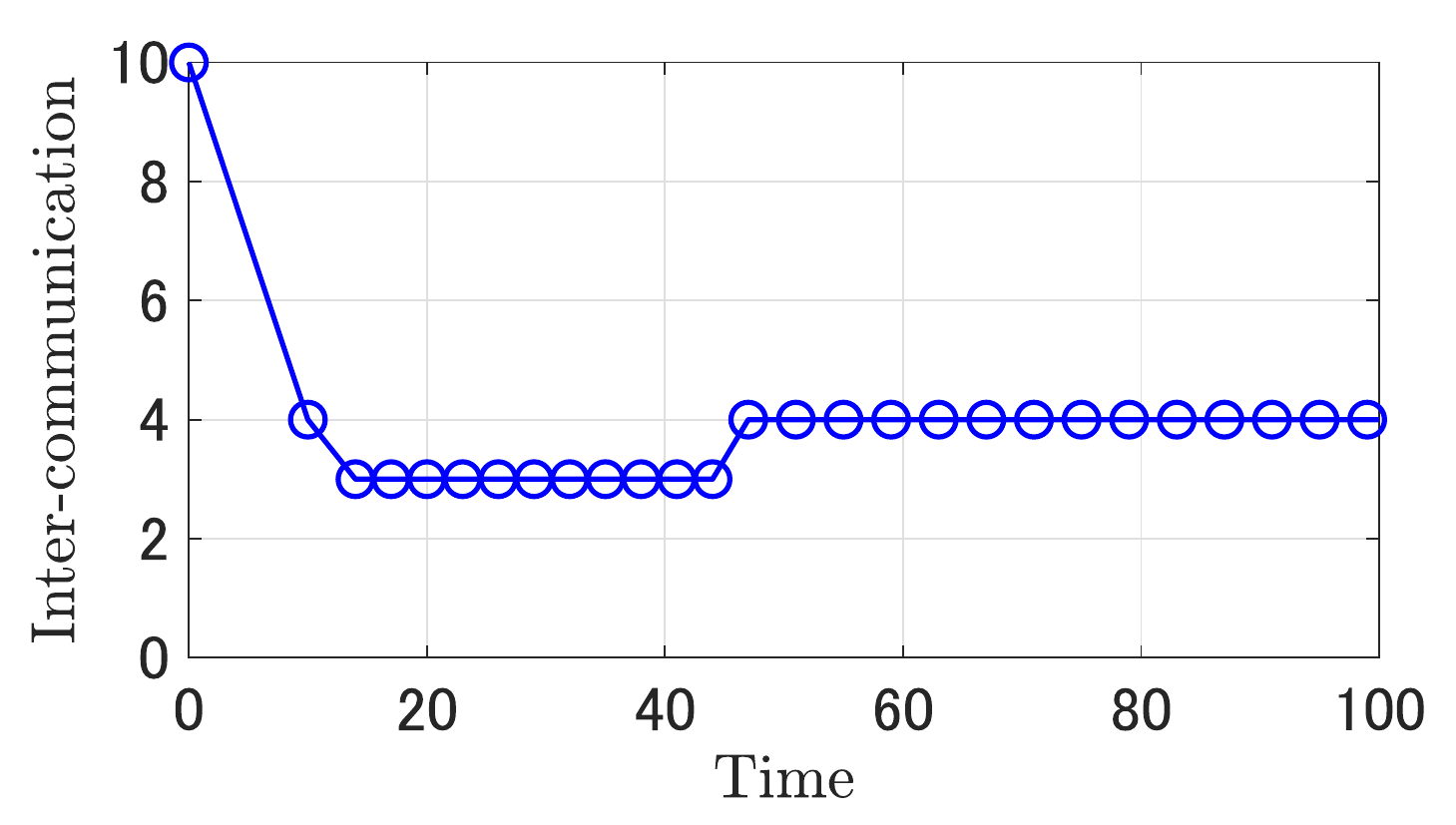}} \label{intercomfig}}
     \subfigure[State trajectories by implementing \ralg{online} with $M=1, 10$.]{
      {\includegraphics[width=8.5cm]{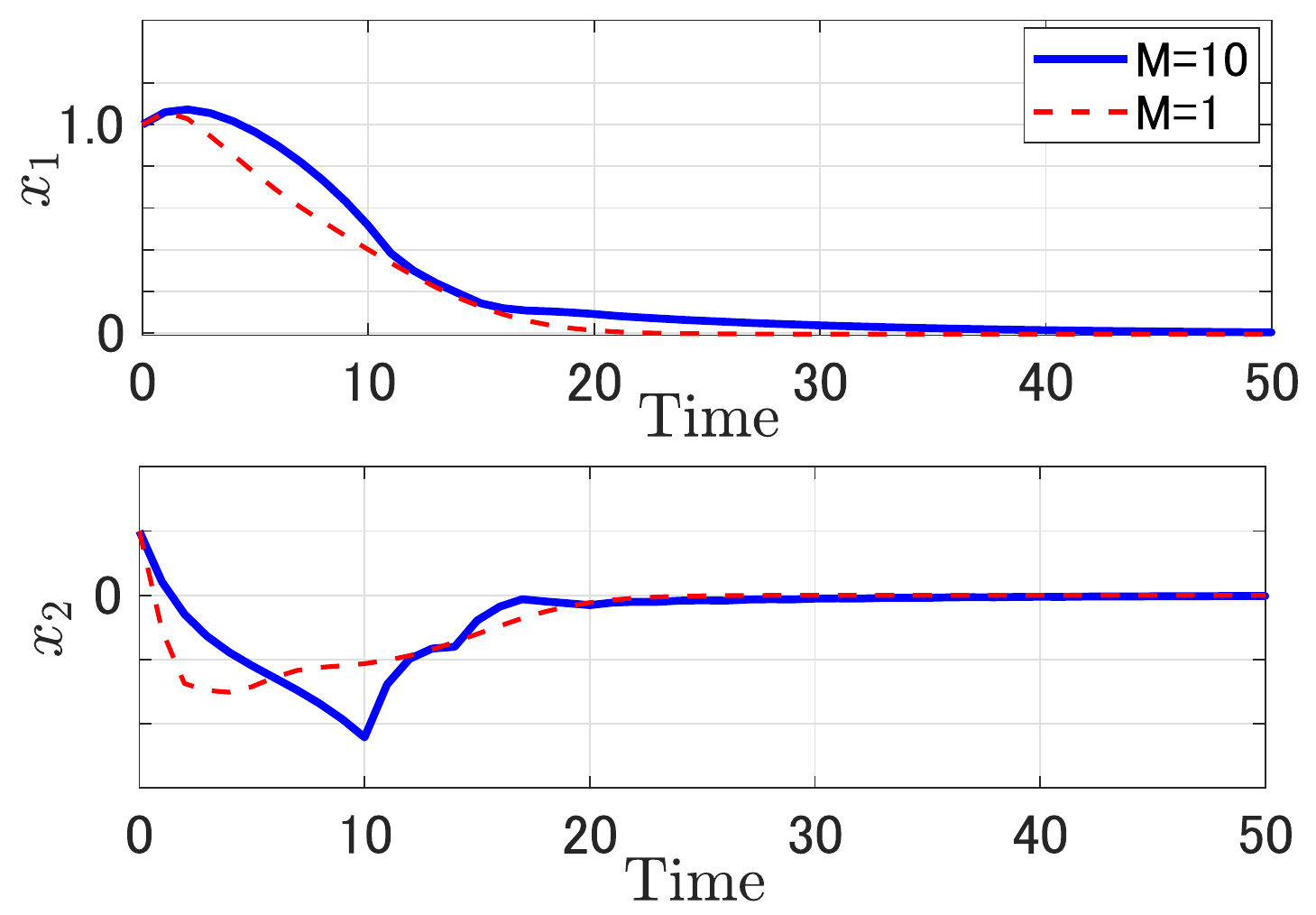}} \label{trajectoryfig2}}
    \caption{{Simulation results by applying \ralg{online}. \rfig{trajectoryfig} illustrates the state trajectories by applying \ralg{online} with $M=10$ after ${\rm Episode} = 1$ (red dotted) and $10$ (blue solid). \rfig{intercomfig} illustrates the corresponding inter-communication time steps for the case ${\rm Episode} = 10$. Moreover, \rfig{trajectoryfig2} illustrates the state trajectories by applying \ralg{online} after ${\rm Episode} = 10$ with different selections of $M$ ($M=1, 10$). Note that, $M=1$ corresponds to the case when communication is given at every time step (i.e., the time-triggered controller is implemented).}}
    \vspace{-0.3cm}
\end{figure}



\begin{figure}[t]
  \begin{center}
   \includegraphics[width=9.0cm]{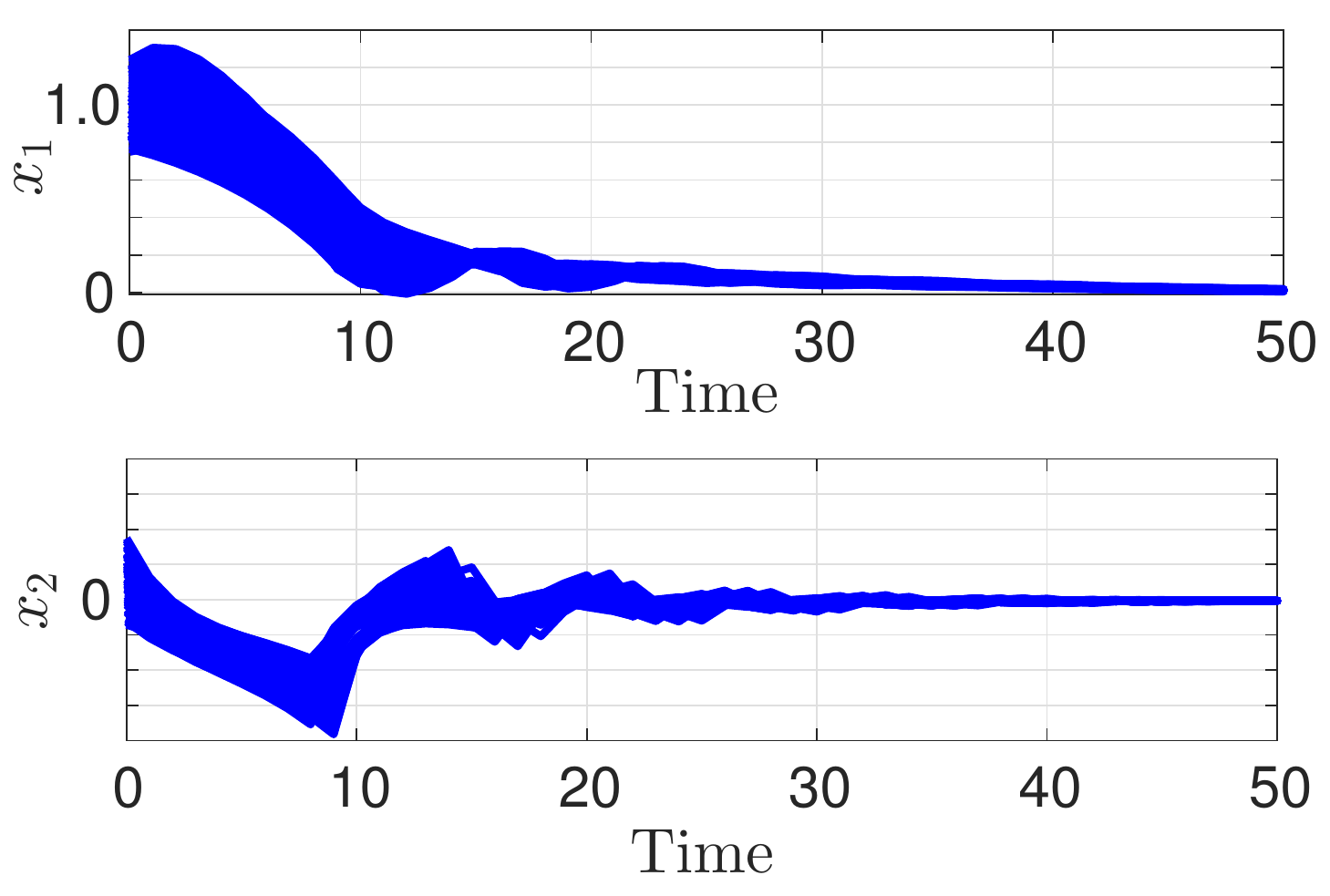}
   \caption{State trajectories from random initial states by applying the derived self-triggered controller.} 
   \label{diftrajectory}
  \end{center}
\end{figure}

\rfig{trajectoryfig} illustrates the trajectories of the states by applying the self-triggered controller obtained by \ralg{online} with ${\rm Episode} = 1$ (red dotted) and $10$ (blue solid). The figures illustrate that, while the state diverges at the initial learning phase, it is indeed stabilized towards the origin as the number of episode increases. The computed inter-communication time steps corresponding to the simulation result in \rfig{trajectoryfig} (${\rm Episode} = 10$) are illustrated in \rfig{intercomfig}, which shows that the communication is given aperiodically according to the derived self-triggered controller. 
{\rfig{trajectoryfig2} illustrates the state trajectories by applying \ralg{online} after ${\rm Episode} = 10$ with different selections of $M$ ($M=1, 10$). Note that, $M=1$ corresponds to the case when communication is given at every time step, i.e., the time-triggered controller is implemented. 
The figure shows that the convergence of states for the case $M=1$ seems to be faster than for the case $M=10$, which is due to the fact that control inputs are updated at every time step when the time-triggered controller is implemented. On the other hand, the total number of communication instants required for the time interval $k\in [0, 100)$ is $100$ for the case $M=1$ (as it is the time-triggered implementation), while it is $27$ for the case $M=10$. This implies that employing the self-triggered controller achieves a significant communication reduction in contrast to the time-triggered strategy. Hence, the result shows that there exists a tradeoff between the communication reduction for the NCS and the convergence speed of states towards the origin, and such tradeoff may be regulated by tuning the parameter $M$.}

To indicate the robustness of the derived self-triggered controller, we also illustrate in \rfig{diftrajectory} several trajectories from different initial states around ${\bf x}_{\rm init}$. The figure illustrates that the states are indeed stabilized to the origin regardless of the deviation of the initial states, showing the robustness of the self-triggered controller. 


\begin{figure}[t]
   \centering
    \subfigure[State trajectories with $\gamma = 0, 0.01, 0.03$.]{
      {\includegraphics[width=8.5cm]{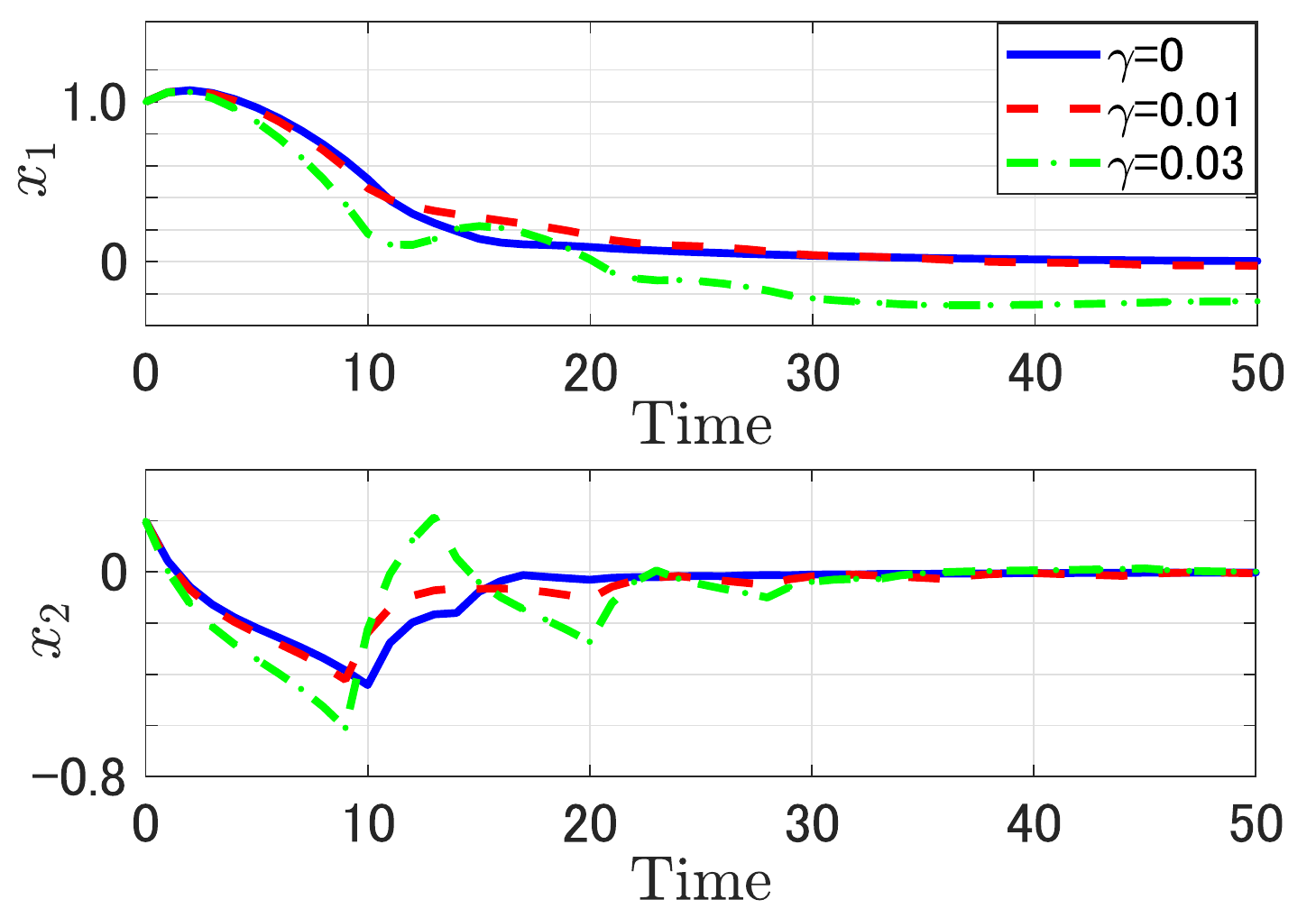}} \label{difgammatrajectory}}
    \subfigure[Inter-communication time steps with $\gamma = 0, 0.01, 0.03$.]{
      {\includegraphics[width=8.5cm]{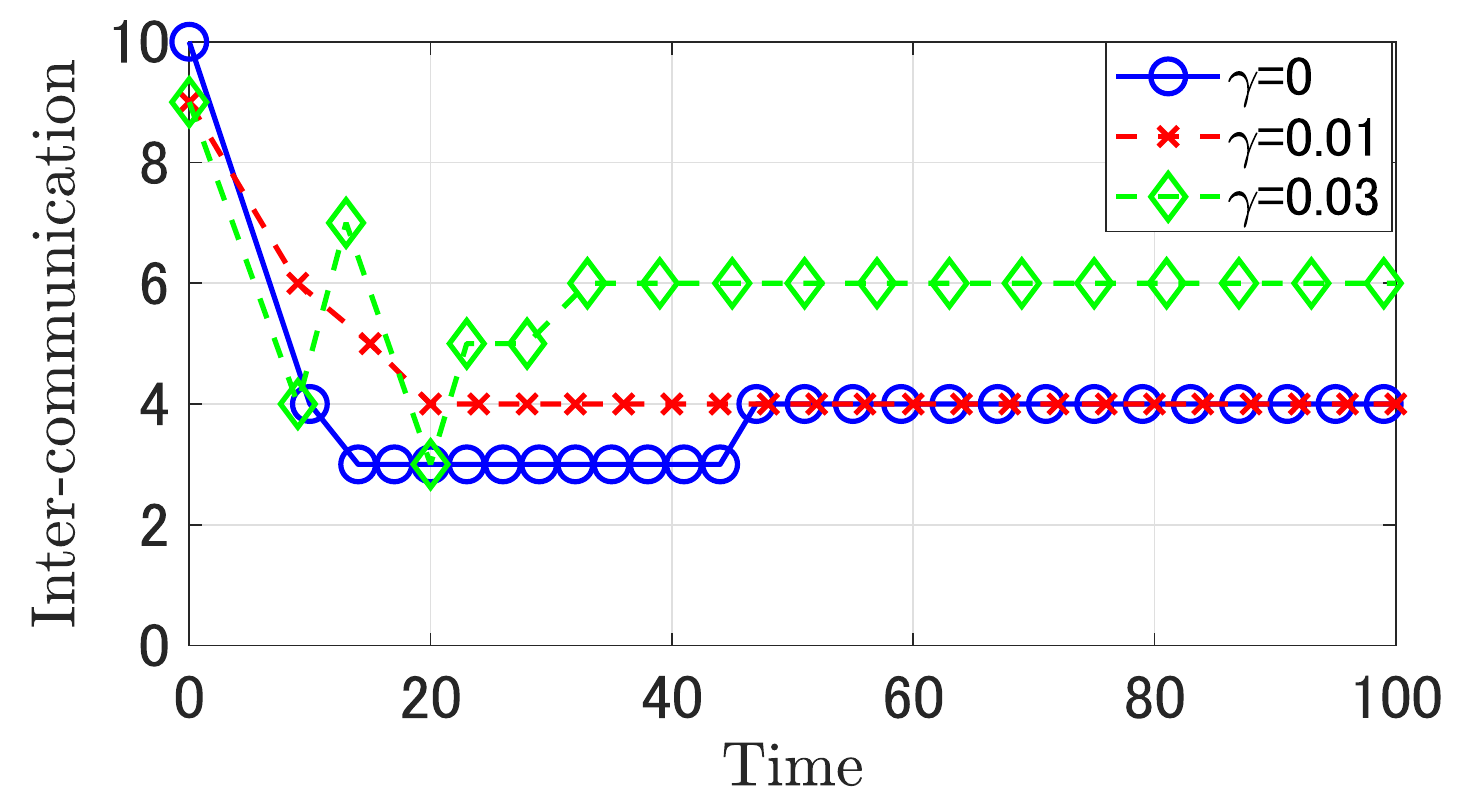}} \label{intercomfig2}}
    \caption{Simulation results with different selections of $\gamma$.}
    \vspace{-0cm}
\end{figure}

Finally, to analyze the effect of $\gamma$, we illustrate in \rfig{difgammatrajectory} and \rfig{intercomfig2} the resulting state trajectories under different selections of $\gamma$ ($\gamma = 0.01, 0.02, 0.03$), and the corresponding inter-communication time steps, respectively. Here, \ralg{online} has been implemented for each $\gamma$ with $10$ episodes ($N_{\rm epi} = 10$). From \rfig{intercomfig2}, it is shown that larger inter-communication time steps are more likely to be selected as $\gamma$ is selected larger. This is due to the fact that, by selecting larger $\gamma$, it will penalize more for the communication cost. Note that, for the case $\gamma=0.03$, the resulting state trajectory converges farther from the origin than for the other cases (while it achieves larger inter-communication time steps), which may be due to the fact that achieving large inter-communication time steps is too emphasized. Hence, similarly to the above, there exists a tradeoff between the communication reduction for the NCS and the convergence of states towards the origin, and such tradeoff may be regulated by tuning the parameter $\gamma$.

\section{Conclusion and future work}
In this paper, we investigate the self-triggered controller for NCSs with the unknown transition dynamics. To this end, we use the GP to learn the dynamics of the plant. 
We first formulate an optimal control problem, such that both the cost for the control performance and the communication cost can be taken into account. Then, we illustrate that the optimal control problem can be solved via a value iteration algorithm, in which the optimal pair of the control input and the inter-communication time steps can be determined based on the GP model of the plant. Then, we provide overall reinforcement learning algorithm that jointly learns the dynamics of the plant as well as the self-triggered controller implemented by the learning agent. Finally, a numerical simulation is given to illustrate the effectiveness of the proposed approach.  

{Future work involves analyzing some theoretical issues (e.g., stability of the closed loop system, convergence property of the value iteration algorithm, etc.) for the GP dynamics of the plant.}
{Moreover, providing some experiments to test the applicability of our approach should be investigated for our future work of research.}

\end{document}